\documentstyle[12pt]{article}
\begin {document}
\title{Formation of a disk-structure and jets in the symbiotic prototype Z And during its 2006-2010 active phase}

\author{T. N.~Tarasova, A.~Skopal}

\maketitle

\begin{abstract}
We present an analysis of spectrophotometric observations of the latest cycle of activity of the symbiotic binary Z And from 2006 to 2010. We estimate the temperature of the hot component of Z And to be~$\approx$~150000--170000~K at minimum brightness, decreasing to $\approx$ 90000~K at the brightness maximum. Our estimate of the electron density in the gaseous nebula is N$_{e}$~=~10$^{10}$~--~10$^{12}$~cm$^{-3}$ in the region of formation of lines of neutral helium and 10$^6$~--~10$^7$~cm$^{-3}$ in the region of formation of the [OIII] and [NeIII] nebular lines. A trend for the gas density derived from helium lines to increase and the gas density derived from [OIII] and [NeIII] lines to simultaneously decrease with increasing brightness of the system was observed. Our estimates show that the ratios of the theoretical and observed fluxes in the [OIII] and [NeIII] lines agree best when the O/Ne ratio is similar to its value for planetary nebulae. The model spectral energy distribution showed that, in addition to a cool component and gaseous nebula, a relatively cool pseudophotosphere (5250--11 500 K) is present in the system. The simultaneous presence of a relatively cool pseudophotosphere and high-ionization spectral lines is probably related to a disk-like structure of the pseudophotosphere. The pseudophotosphere formed very rapidly--over several weeks--during a period of increasing brightness of Z And. We infer that in 2009, as in 2006, the activity of the system was accompanied by a collimated bipolar ejection of matter. In contrast to the situation in 2006, the jets were detected even before the system reached its maximum brightness. Moreover, components with velocities close to 1200 km/s disappeared at the maximum, while those with velocities close to 1800 km/s appeared.
\end{abstract}

\maketitle
\section{Introduction}
Z And is one of the most active symbiotic stars. After extensive spectrophotometric observations of Z And from 1960 to 1965 \cite{B}, it became commonly accepted that the system is a binary with a cool giant and hot dwarf component, surrounded by a nebula. Studies from the ultraviolet to the infrared \cite{FC} confirmed the binarity of this system. Murset and Schmid \cite{MS} have shown that the giant has spectral type M4.5. It was also established that the accretion of the stellar wind of the giant by a 0.5--1~M$_{\odot}$ white dwarf results in the formation of a very high temperature (up to T$\approx$10$^{5}$~K) and high luminosity (up to 10$^4$\,L$_{\odot}$) emission source that ionizes part of the gaseous envelope of the system and produces the nebular emission spectrum. Using photometric observations covering 112 years, Leibowitz and Formiggini \cite{LF} discovered that the activity of the system is quasi-periodic. The period of the activity is 15--25 years. Active phases were observed in 1915, 1939, 1960, 1985, and 2000. In the active phase, a series of outbursts occurs, during which the brightness of Z And increases by 1.5$^m$ (sometimes, by 2$^m$--3$^m$), then decreases to its minimum level. 

Currently, models are suggested, in which the increased activity of Z And is related to an increase in the accretion rate onto white dwarf. The necessary accretion rate can be provided by disruption of the accretion disk. Such disruption is possible \cite{BBKK}, and even a small change in the velocity of the wind from the cool component is sufficient for this. As a result, an optically thick envelope forms on the surface of white dwarf. Thermonuclear reactions at the base of the envelope result in the envelope's expansion and, accordingly, in brightening of the system. 

During the previous outburst, in 2000, a gradual brightening of the system was observed. Bisikalo et al. \cite{BBKTT} and Sokoloski et al. \cite{SKE} carried out a detailed analysis of particular stages of the outburst, and showed that these may be considered to be due to the manifestations and interactions of two mechanisms. The first is disruption of the disk, and the second is the presence of thermonuclear reactions at the whitedwarf surface. An additional supply of matter from the accretion disk provides the accretion rate required for thermonuclear reactions, which explain the observed high luminosity of the hot component, up to L$_{hot}\approx 10^{4}$~L$_{\odot}$. However, although the existence of a slowly expanding disk-like structure in the initial stage of the outburst was noted by Skopal at al. \cite{SVE}, who studied the active phases in 2000--2003, direct evidence for the existence of a disk-like accreting object was absent before the discovery of transient jet-like structures in the radio \cite{BSK} in 2001 and the optical in 2006 \cite{SP,SPB}. The existence of jets was later confirmed by Burmeister and Leedjarv \cite{BL}, and Tomov et al. \cite{TTB}. All these observational data provide direct evidence for the existence of an accretion disk in the system and its role in the outburst. 

Z And entered an active phase in 2000. Six outbursts have been observed since that time. Three of these were analyzed by Tomov et al. \cite{TTB2010}. The last outburst, at the end of September 2009, was the most powerful--the visual brightness of the star increased by more than 3$^m$. In the current paper, we analyze the behavior of the system in its active state and in quiescence, using photometric and spectral observations carried out during periods of activity of Z And in 2006, 2008, and 2009--2010 and in periods of quiescence in 2007 and 2009.

\section{Observations}

The spectral observations were obtained on 2.6-m Shain mirror telescope (ZTS) with low (700) and medium (1300) resolution. A journal of the observations is presented in Table~1. The spectra were obtained using a slit spectrograph (SPEM) placed at the Nasmyth focus. The detector was a SPEC-10 1340$\times$100 pixel CCD. With 651 lines/mm, the dispersion was close to 2 \AA/pixel (resolution 700), while the dispersion was close to 0.75~\AA/pixel with 1200 lines/mm (resolution 1300). Two to five images were obtained during a night, for two or three spectral regions centered either on the wavelengths of H$_{\alpha}$ and  H$_{\beta}$ or on the central wavelength of the reaction curves for the Johnson B and V bands. A total low-resolution spectrum was obtained by fitting the partially overlapping spectra. Table 1 gives the Julian date of the observations phase of the orbital motion and spectral region. The phase of the orbital motion was calculated using the ephemeris of Formiggini and Leibowitz \cite{FL1994}: Min(vis)~=~JD2442666($\pm$10) + 758.8($\pm$2)E. 

The preliminary reduction of the spectra, which included bias subtraction and correction for inhomogeneity of the sensitivity of the CCD field, was carried out using the SPERED code written by S.I. Sergeyev at the Crimean Astrophysical Observatory. The wavelength scale was calibrated using a HeNe lamp. The fluxes in the stellar spectrum were calibrated using the absolute spectral energy distribution (SED) of the spectrophotometric standard HR 8965 from the catalog \cite{Burn}. This spectrophotometric standard was observed on the same date and at the same zenith distance as Z And. Therefore, we did not take into account the difference in the air masses for the standard and Z And. Since we used a slit spectrograph, we compared the magnitudes of Z And in the B and V filters to the B and V magnitudes calculated based on calibrated spectra obtained on the same dates as a check. If there were no magnitudes for a given date, we calculated them by interpolating between the magnitudes obtained on nearby dates. The differences between the calculated and measured magnitudes were, on average, close to 0.1m. The flux calculations took into account reddening of the star due to interstellar absorption. The color excess was taken to be E(B-V)~=~0.3~\cite{MK1996}. The distance to the star was taken to be 1.5 kpc \cite{MK1996}. 

Classical photometric observations in the UBVR bands were carried out using a one-channel photometer placed at the focus of the 0.6-m reflector of the Skalnate Pleso Observatory. The star BD + 47 4192 (SAO 53150, V = 8.99, B--V = 0.41, U--B = 0.14, V--R$_C$~=~0.10) was adopted as the standard. The internal daily average errors did not exceed several thousandths of a magnitude. The light curve of Z And in 2006--2010 obtained from these data is shown in Fig. 1.

\begin{center}
Fig 1. Light curve of Z And in 2006--2010.
\end{center}

\section{Spectrophotometric evolution}
Spectrophotometric observations of Z And were carried out from September 2006 to November 2010. During this period Z~And experienced three outbursts. In July 2006, the Nova increased its visual brightness by 2.5$^m$, reaching its historical maximum of 8$^m$. In July 2008, the observed brightness increased by 1.5$^m$, and the brightness of the star increased by more than 3$^m$ over a month in September 2009, surpassing the 2006 maximum. Figure 1 shows a light curve of Z And for this period, based on data from the database of A. Skopal and new unpublished data for 2009--2010. Figure 1 shows variations of the star's brightness in the U, B, V and R bands of the Johnson system during the period of our observations. The vertical marks in the light curve indicate the dates of our spectral observations.  The spectra are shown in Fig. 2--6. We measured the fluxes for all spectral lines that were confidently detected in the spectra, presented in Table 2. The errors in the fluxes for strong lines were close to 2\%--5\%, while they were 10\%--20\% for weak lines. Below, we present a brief description of the active phases in 2006, 2008 and 2009-2010, and  the quiescence phases in 2007 and 2009. 

\subsection{Activity Phase in 2006--2007}
The spectra in this phase of activity (Fig.2) were obtained with the orbital phases of the system 0.93, 0.96 and 0.087. Phases corresponding to the giant being in the front of the white dwarf. Emission lines of neutral hydrogen HI and helium HeI, ionized helium HeII~4686~\AA, and the blend NIII~(4643,4641,4642)~+~CIII~(4647,4650)~\AA\, are prominent in the spectra. In addition, the spectrum has fairly strong forbidden lines: [OIII] 4363, 5007~\AA, [NeIII] 3869~\AA\, and [NeIII] 3968~\AA\, and weak lines of ionized iron FeII.

\begin{center}
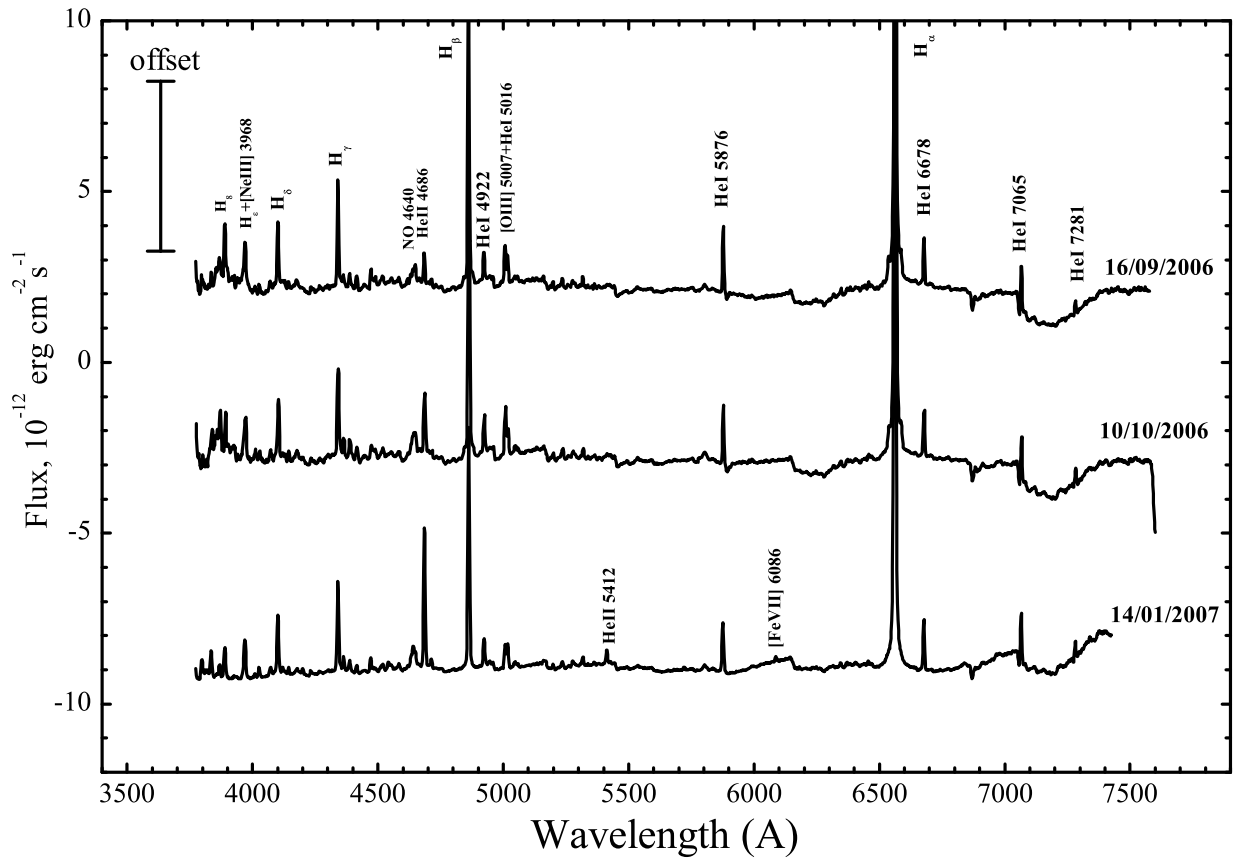
Fig 2. Low-resolution spectra obtained in the activity phase in 2006--2007. The fluxes are displaced relative to each other by a fixed value, starting from the first spectrum. The numbers to the right of each spectrum give the date of the observations.
\end{center}

On September 16, 2006 and October 10, 2006, features on the blue and red sides of the bases of H$_{\alpha}$ and H$_{\beta}$ were present (Fig. 7). We were able to distinguish these features in line blends at the wavelengths of H$_{\alpha}$ and H$_{\beta}$. They indicate that we were observing collimated ejections of matter with radial velocities close to 1200 km/s. On dates close to those noted above, Skopal et al. \cite{SP,SPB}, Burmeister and Leedjarv \cite{BL}, and Tomov et al. \cite{TTB} observed satellite lines displaced to the blue and red sides of the main components of H$_{\alpha}$ and H$_{\beta}$ with essentially the same velocities. 

\subsection{Quiescent Phase in 2007}
In 2007, we obtained two spectra (Fig. 3) when Z And was in an inactive state, with the phases of the orbital motion (0.35 and 0.43) corresponding to the white dwarf being in the front of the giant. The flux in the HeII 4686 line increased considerably (by a factor two) after the active phase. As in the previous phase of activity decline, a decrease of the flux in all HeI lines was observed, apparently also related to an increase of the helium ionization. The appearance of [FeVII] 3587, 3760, 5159, 5721, 6086~\AA\, spectral lines, as well as the Raman scattering 6830~\AA\, indicate the generation of high-temperature radiation. Along with emission lines of elements with high-ionization potentials, molecular lines of TiO belonging to the giant are also present.

\begin{center}
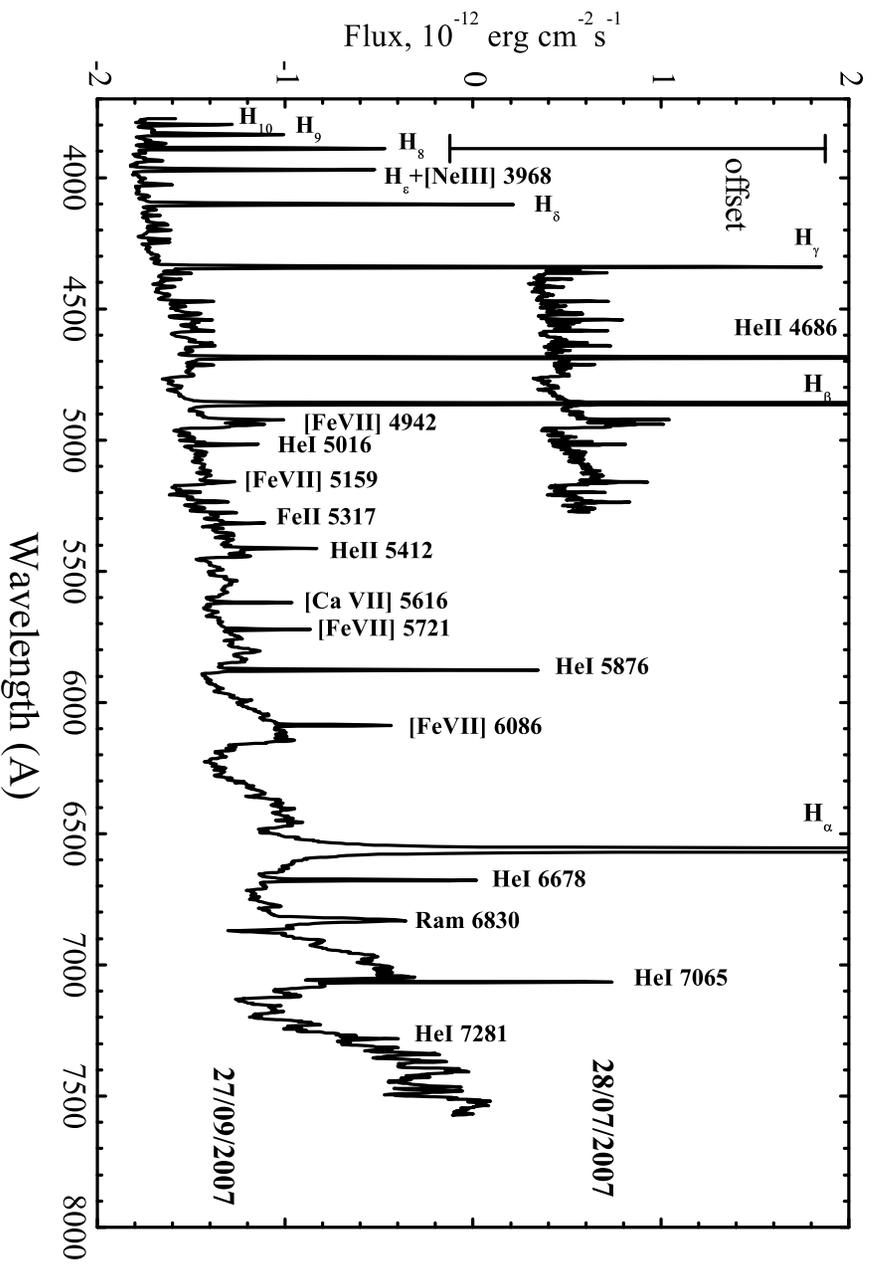
Fig 3. Low- and medium-resolution spectra obtained in the quiescent phase in 2007.
\end{center}

\subsection{Activity Phase in 2008}
In 2008, we observed the system when it resumed its activity and changed its visual brightness by 1.5$^m$. We obtained 11 spectra during this phase (Fig. 4). The first two were obtained at brightness maximum, and the others during a small plateau in the light curve. The last two spectra were observed as the brightness of the star declined. The phase of these observations, 0.8--0.96, corresponds to the stage when the white dwarf and part of the gaseous nebula are eclipsed by the giant. The most prominent variations in this phase are the decline of the [FeVII] flux by a factor of four to five, the disappearance of the Raman scattering line 6830~\AA, and an enhancement of the fluxes in the [OIII] 4363~\AA\, line (by a factor six) and [OIII] 5007~\AA\, line (by a factor of ten), compared to the quiescent phase in 2007. Our model SED (Fig. 11) shows that, in addition to the giant and nebula, a fairly cool component is present in the system (with a temperature of 5750 K), which we refer to as the pseudophotosphere.

\begin{center}
Fig 4. Low- and medium-resolution spectra obtained in the activity phase in 2008.
\end{center}

\subsection{Quiescent Phase in 2009}
The phase of the orbital motion in 2009 varied from 0.23 to 0.35; i.e., the eclipse of the white dwarf was over. The high ionization potential lines [FeVII], [NeV], [CaVII], [NeIII], OIII 3443~\AA\, again appeared in the spectrum (Fig.5). In contrast to the 2007 minimum, there is no Raman scattering line in the spectrum, despite the fact that the quiescent phases in 2007 and at our latest date in 2009 differ only slightly, and the phases are such that the white dwarf is not eclipsed by the giant. Therefore, we suggest that the absence of flux from the Raman scattering line is due to the appearance of an additional source eclipsing the region close to the white dwarf. Our model SED (Figs. 12--14) shows that an additional source of emission (a pseudophotosphere) is indeed present in the spectrum, athough its contribution is not too large. The Balmer jump is clearly visible in all the spectra.

\begin{center}
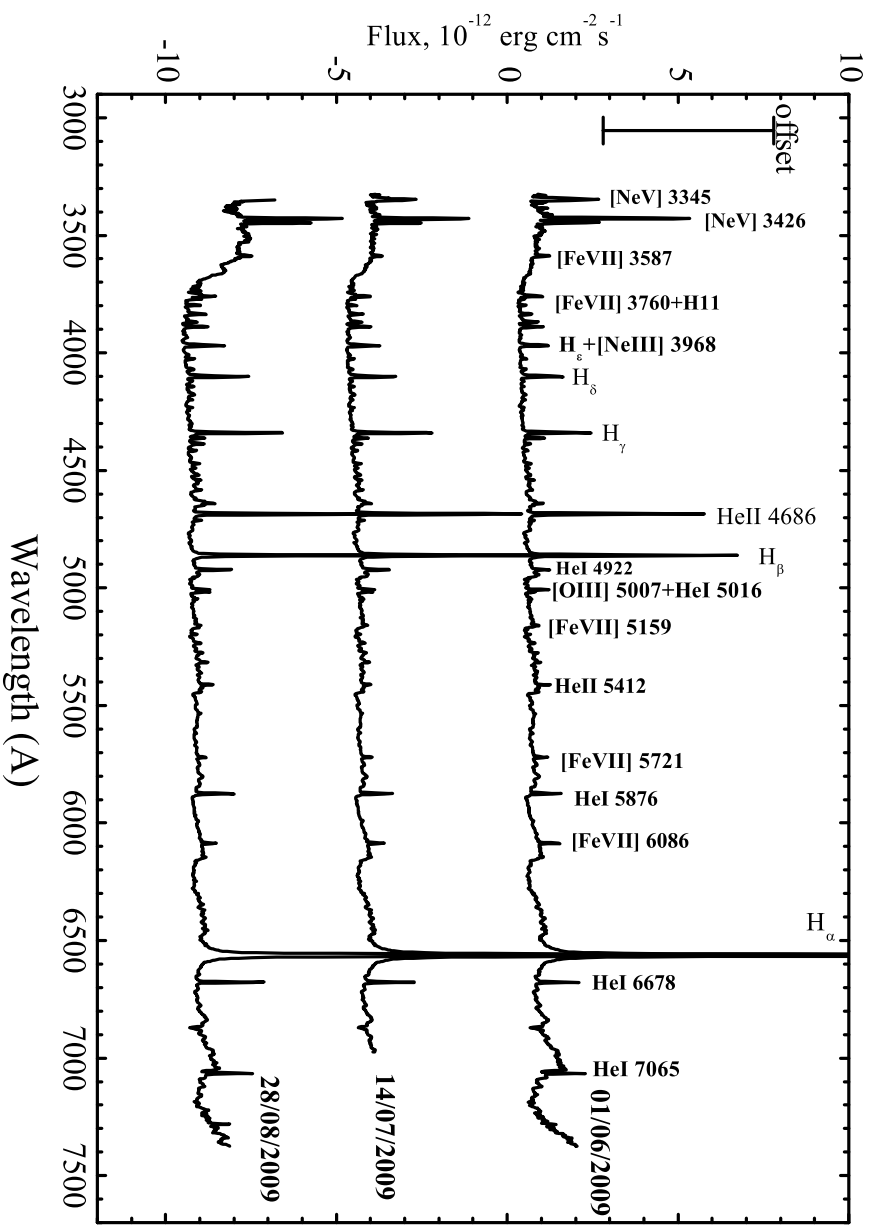
Fig 5. Low-resolution spectra obtained in the quiescent phase in 2009.
\end{center}

\subsection{Activity Phase in 2009--2010}
During this period, the star was again very active, and the visual magnitude of the system reached 8$^m$. The 2009 observations (fig. 6) were made at the orbital phase 0.39, when the white dwarf was not eclipsed by the giant. In spite of this fact, we detected lower fluxes in HI and HeI lines in this phase, compared to the fluxes in the previous activity periods in 2006 and 2008. The [FeVII], [NeV], [CaVII] lines were significantly weakened (by almost an order of magnitude) or had disappeared completely. As before, there is no 6830~\AA\, Raman scattering line. The fluxes in the [NeIII] 3869~\AA , [OIII] 4363~\AA\, lines remained nearly constant. Only the [OIII] 5007~\AA\, flux increased by a factor 2.5--3, becaming similar to its fluxes in the previous phases of activity of Z And. There is no Balmer jump in the spectrum. The model SED (Figs. 15,16) showed that the disappearance of the Balmer jump is related to the appearance of an additional radiation source that provides a significant contribution in this region, exceeding that of the nebula. The 2010 observations correspond to orbital phases 0.7--0.85. In 2009--2010, when the brightness of the system increased, i.e., it passed into a stage of activity, we observed features on the blue and red sides of the H$_{\alpha}$ and H$_{\beta}$ wings, similar to those observed in 2006 (Fig. 7). In the phase of brightness increase, these features had the same radial velocities as in 2006, close to 1200 km/s. However, after the star reached its maximum brightness, new components appeared, with a significantly higher velocity-up to 1800 km/s. The existence of these features was noted earlier by Skopal et al. \cite{STP}, and was confirmed by Tomov et al. \cite{TTB2010}. The H$_{\alpha}$ and H$_{\beta}$ wings are presented in Fig.~7. The arrows show the positions of the satellites of the H$_{\alpha}$ and H$_{\beta}$ lines.

\begin{center}
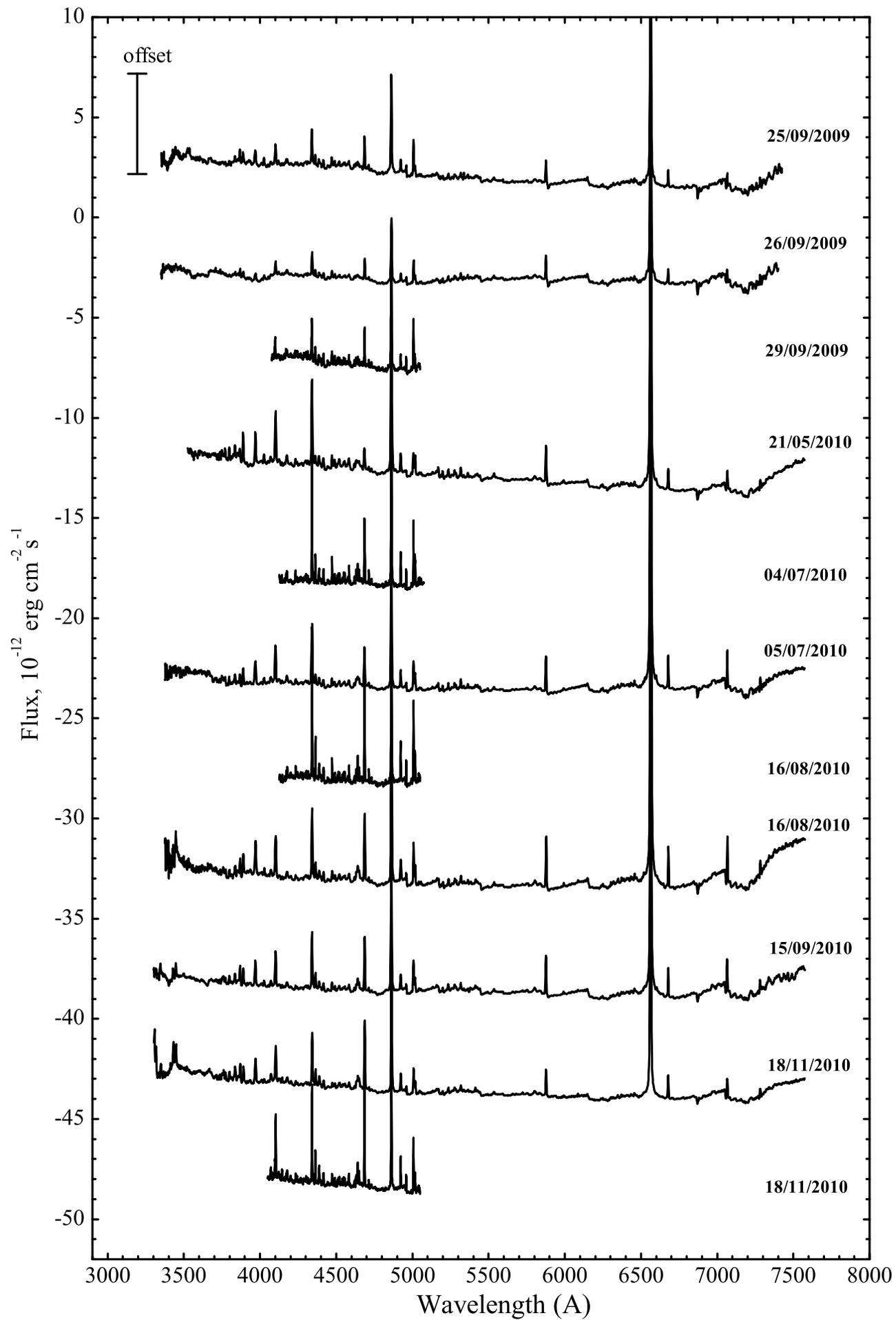
Fig 6. Low- and medium-resolution spectra obtained in the activity phase in 2009--2010.
\end{center}

\begin{center}
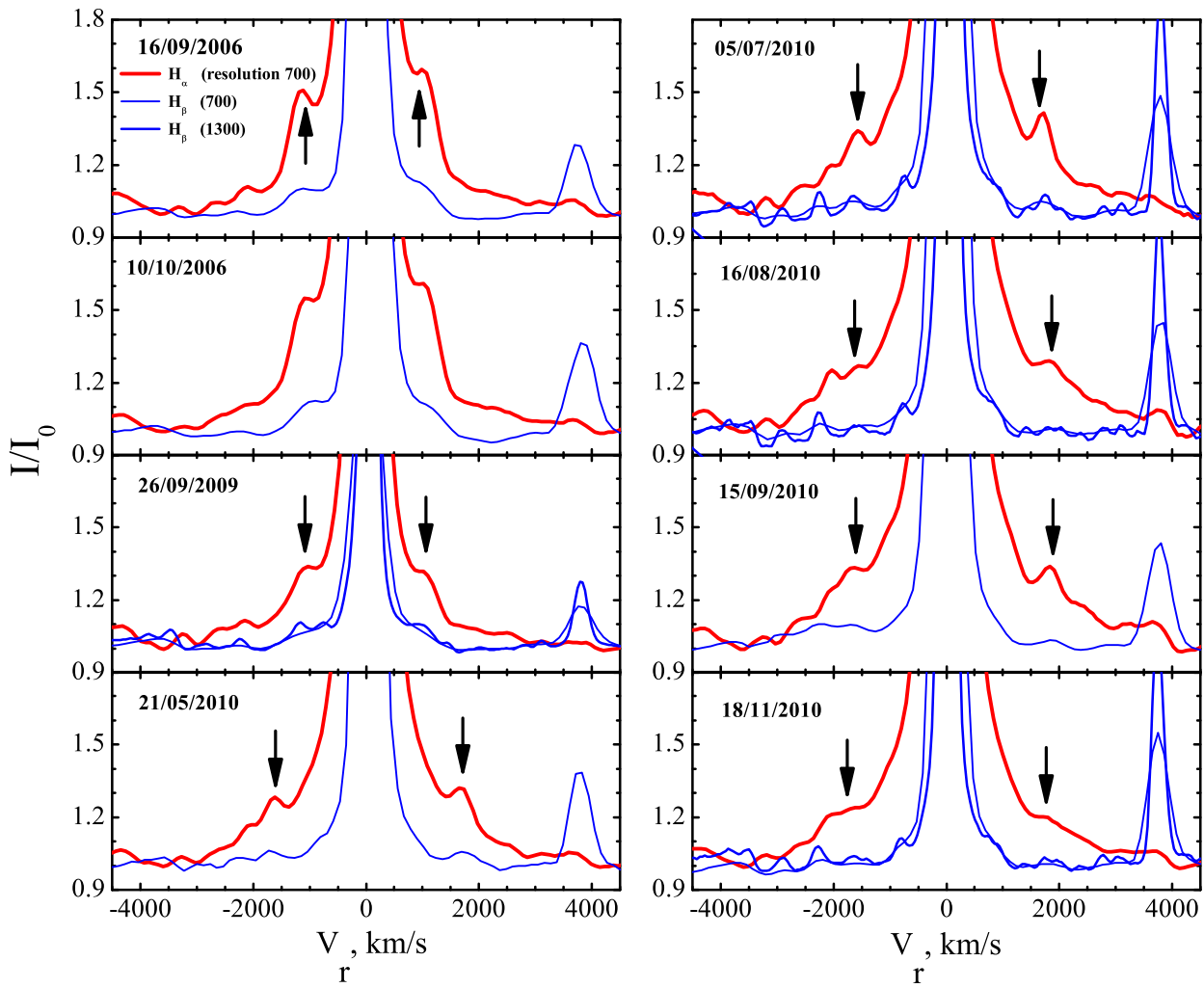
Fig 7. Profiles of the H$_{\alpha}$ and H$_{\beta}$ wings indicating collimated ejections of matter in Z And.
\end{center}

\subsection{Results}
Figure 8 shows the variability of the fluxes in the strongest lines of elements with various ionization potentials. For clarity, light curves derived from V observations are shown in the upper right and left plots. Our analysis of the variations of the line fluxes led to the following results.

\begin{center}
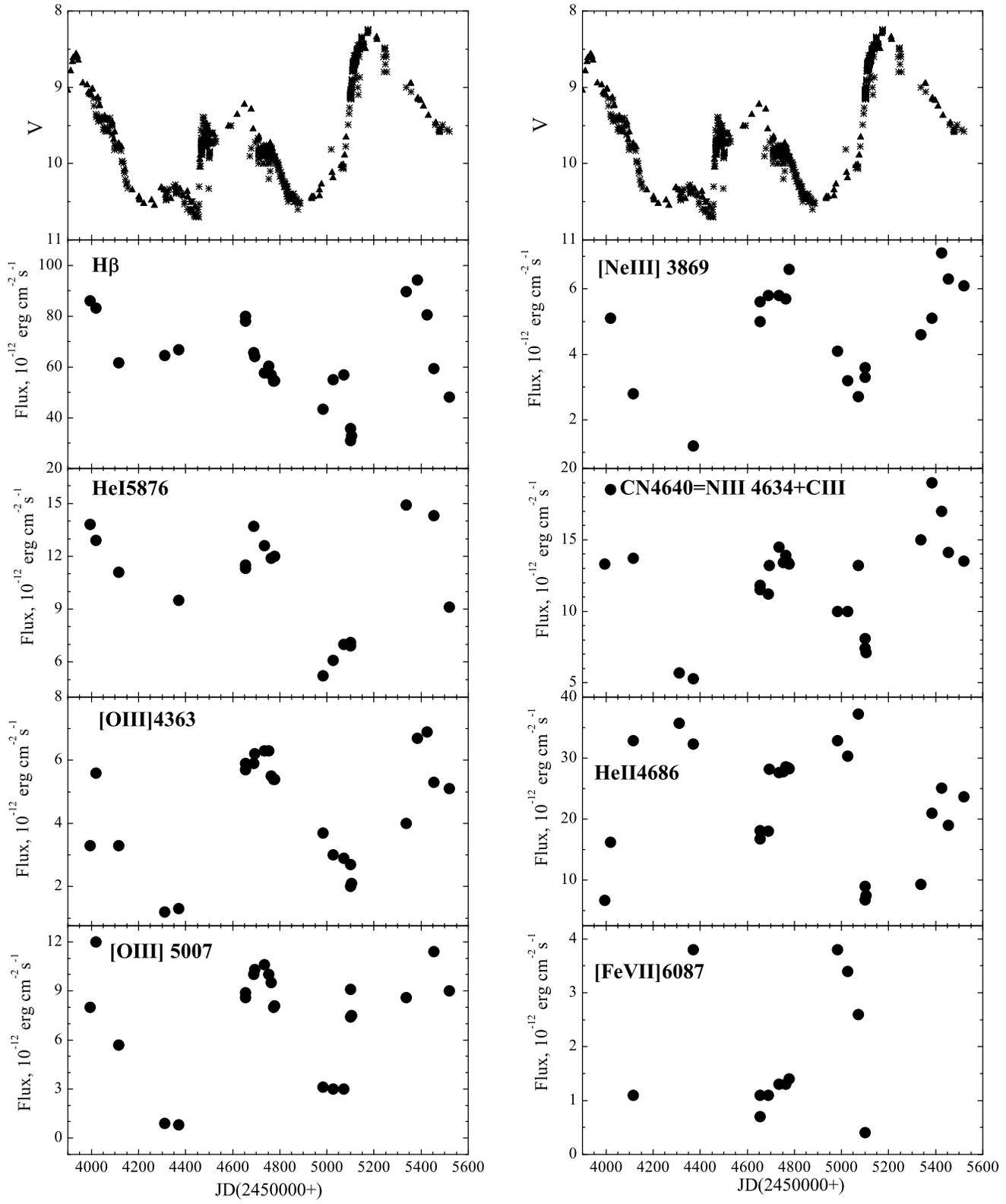
Fig 8. Variations of fluxes in lines of elements with various ionization potentials. The fluxes of CN~4640~\AA\, were determined as the fluxes of the blend formed by the NIII 4643, 4641, 4642~\AA\, and CIII 4647~\AA\, lines.
\end{center}

1. The HI,  HeI,  [OIII], [NeIII] and NIII 4640 line fluxes increased when the system was active and decreased when it was quiescent. However, at the phase close to the maximum brightness in 2009, a considerable weakening of the fluxes in almost all spectral lines was observed. The flux apparently decreases at maximum brightness because the lines are veiled by an enhanced continuum produced by an optically thick envelope (pseudophotosphere) that formed around the white dwarf in the phase of maximum brightness.

2. On the contrary, in contrast to the HI and HeI lines, the fluxes in the HeII and the [FeVII] and [NeV] line were enhanced in quiescence and reduced in active states. In addition, the fluxes in the HeII lines were the same during the quiescent states of the system in 2007 and 2009. On the other hand, the fluxes in the active phase in 2008 were higher than those at the brightness maximum during the phase of activity in 2009. This seems to provide evidence that these outbursts have different characters (different temperatures of the pseudophotosphere).

3. The 6830~\AA\, Raman scattering line is present in the spectra taken during the minimum in 2007, but absent during the minimum of 2009, despite the fact that the orbital phases were very similar, corresponding to the uneclipsed white dwarf. We suggest that the weakening of the flux during the 2009 minimum is associated with the appearance of an object that eclipses the region close to the white dwarf, which is the source of high-energy radiation responsible for the formation of the 6830~\AA\, scattering line. Indeed, our model SED (Figs.~13,~14) indicates the contribution in the spectrum of a cool pseudophotosphere.

\section{Hot component}
We estimated the temperature of the hot component using the expression from \cite{I}:
The results are presented in Table 3. Our estimates for 2006 are very close to those of Burmeister and Leedjarv \cite{BL}. Close to the brightness maximum in 2009, the temperature of the hot source, 118 000--112 000 K, was somewhat higher than the temperature estimates 76 000--114 000~K of \cite{BL} and the temperature 90 000--112 000~K estimated for 2006 (Table 3), and was nearly equal to the temperature estimated for 2008. As the brightness declined, the temperature increased to 140 000--150 000~K (despite the fact that the system did not reach a maximum). In quiescence in 2007 and 2009, the temperature of the hot source was 150~000--170~000~K. 

\section{The nebula}

The HeI 5876, 6678, and 7065~\AA\, lines be used to determine the physical parameters of the gaseous nebula. We compared the ratios of the fluxes in these lines derived from our observational data, with the theoretical computations for symbiotic stars of Proga et al. \cite{Prog}. The results are presented in Fig. 9, which shows the dependence of the HeI flux ratios F(7065)/F(5876) and 
F(7065)/F(5876) for the electron number densities N$_e$ = 10$^6$~cm$^3$ (solid), N$_e$ = 10$^{10}$ cm$^3$ (dashed), and N$_e$ = 10$^{12}$ cm$^3$ (dash--dotted) and temperatures T$_e$ = 10 000, 15 000, and 20 000 K (thick, normal and thin curves, respectively). The ratios corresponding to the radiative mechanism \cite{Ost} are shown in Fig. 9 by the open triangle for T$_e$=10 000 K and the filled triangle for T$_e$ = 20 000 K.

\begin{center}
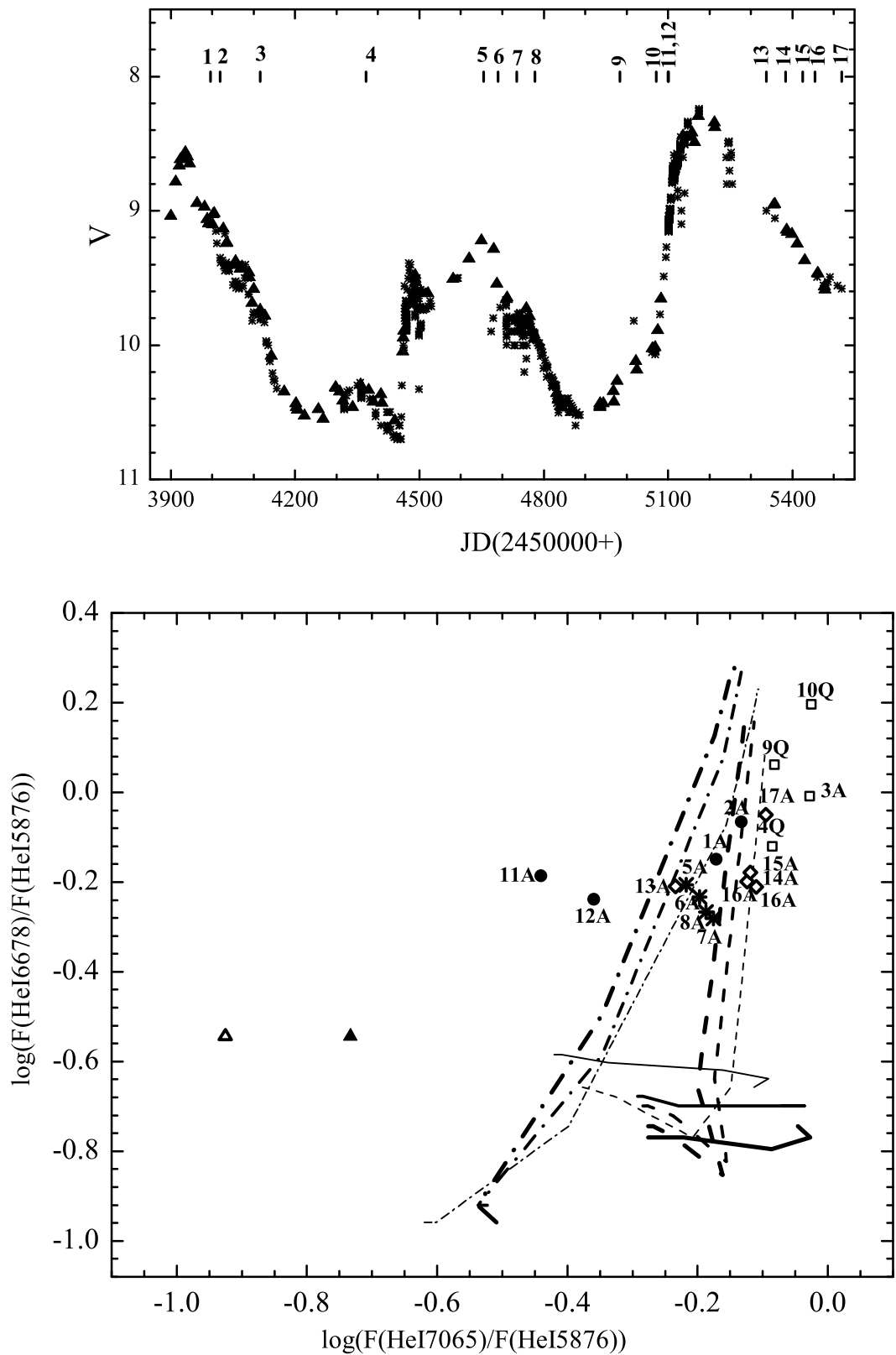
Fig 9. Dependence of the logarithm of the line-flux ratio F(HeI 6678)/F(HeI 5876) on the logarithm of the line-flux ratio F(HeI 7065)/F(HeI 5876) (formore details see Section 5).
\end{center}

Three groups of flux ratios can be distinguished in Fig 9. The ratios of the first group were obtained in quiescence (4Q, 9Q, 10Q) and in stages of brightness decline (1A, 2A, 3A, 14A, 15A, 16A, 17A), of the second group in the active state in 2008 (5A, 6A, 7A, 8A), and of the third preceeding the brightness maximum in 2009 (11A, 12A). 

The helium-line flux ratios for the first group, i.e., following the brightness maximum, are located in the region of lower electron densities, N$_e$ = 10$^{10}$ cm$^3$. In the active state in 2008 (5A, 6A, 7A, 8A), the flux ratios of HeI are located in the region of densities intermediate between N$_e$ = 10$^{10}$ cm$^3$ and N$_e$ = 10$^{12}$~ cm$^3$. The points corresponding to the phase close to the brightness maximum during the 2009 active state occupy a distinct position in the plot, corresponding to densities higher than N$_e$~=~ 10$^{12}$~cm$^3$. Similar flux ratios were found by Skopal et al. \cite{SVE} in 2000, when the brightness of the system was maximum. This suggests that the number density of electrons increases during brightness maxima. 
The HeI 5876~\AA\, line can be used to estimate the size of the region of formation of this line:
\begin{eqnarray}
R_{eff}(He)=\left(\frac{3F_{5876}~d^2}{N_e^2\epsilon_{5876}}\right)^{1/3}
\end{eqnarray}
where $\epsilon$$_{5876}$=10$^{-26}$~erg~cm$^3$~s$^{-1}$  \cite{Prog} for N$_e$~$\approx$~10$^{10}$$\div$10$^{12}\,$cm$^{-3}$  and $\tau_{3889}$~=~100 d~=~1.5~kpc. If the electron number density is N$_e$=10$^{10}$\,cm$^{-3}$, the size of this region is 100 -- 150~R$_{\odot}$ for F(5876)~$\approx$~9$\div$18$\times$10$^{12}$~erg~cm$^{-2}$~s$^{-1}$ while the emission region is very compact if N$_{e}$=10$^{12}$\,cm$^{-3}$~--~5~R$_{\odot}$ \- for F(5876)~$\approx$~5$\times$10$^{12}$~erg~cm$^{-2}$~s$^{-1}$. 
This size is substantially smaller (almost a factor of 40) than the component separation (1 AU). Similar estimates of the density and radius were obtained by Skopal et al. \cite{SVE} for the active phase during the brightness maximum in 2000. 

Thus, at phases close to the maximum brightness, the HeI line emission forms in a fairly compact (5~R$_{\odot}$), dense region (N$_{e}$=10$^{12}$\,cm$^{-3}$), under the condition that this region is homogeneous.

To determine the electron number density and temperature of the nebula, we also used the observed nebular-line flux ratios R$_1$= $\frac{[OIII](4959+5007)}{[OIII](4363)}$ and R$_2$= $\frac{[OIII](4959+5007)}{[NeIII](3869)}$ and compared these to the theoretical values presented in \cite{FerSH}.

Figure 10 (a,b) presents theoretical curves showing variations of R$_1$ and R$_2$ as functions of the N$_e$ and temperature T$_e$. The solid curves with various symbols correspond to various electron number densities, shown in the figure. The temperature varies from 10~000 to 20~000~K in steps of 1000 K along the curves (from right to left). Figures 10(a) and 10(b) show the results for O/Ne abundance ratios of seven and five. The former case corresponds to the chemical composition of the cool component free of anomalies, and the second to the chemical composition of a planetary nebula that is a member of the disk component of the Galaxy \cite{LuCo}. The open triangles show the observed values of R$_1$ and R$_2$ in quiescence and at epochs when the system was in the small plateau of the light curve after the 2008 maximum. The solid circles show the observed values of R$_1$ and R$_2$ corresponding to the active state of the system, when its brightness was close to maximum.

\begin{center}
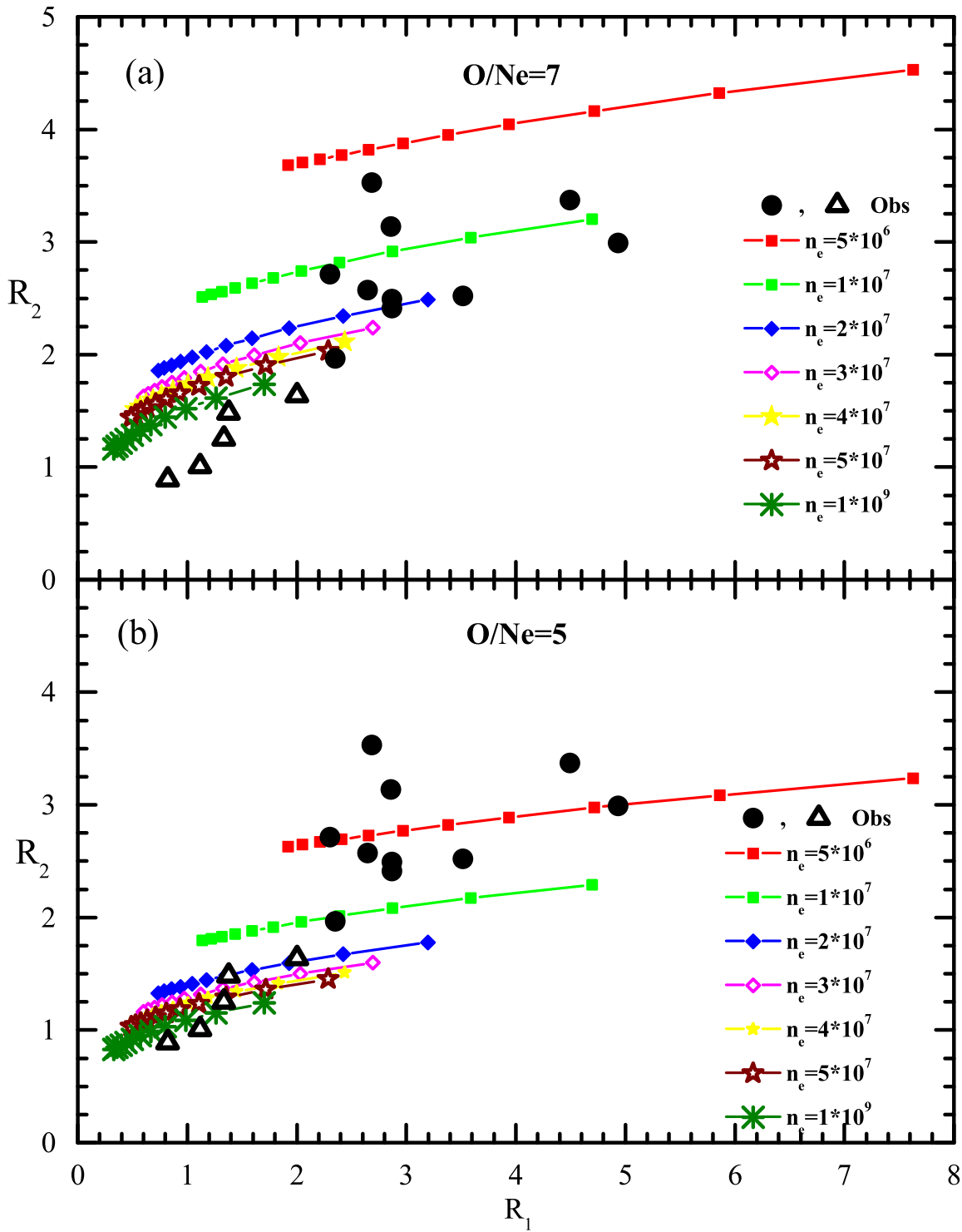
Fig 10. Diagram showing the variation of R$_1$~--~R$_2$ (see the text) for various N$_e$ and T$_e$ values and constant ratios of the oxygen and neon abundances: (a) O/Ne = 7 and (b) O/Ne = 5. The electron number densities corresponding to the various curves are indicated in the plots. Along every curve, the temperature changes from 10 000 to 20 000 K from right to left in steps of
1000 K. The open triangles show the observed values of R$_1$ and R$_2$ corresponding to the quiescent state of Z And. Filled circles correspond to the active state.
\end{center}

Figure 10 shows that the theoretical and observed values of R$_1$ and R$_2$ agree best when the O/Ne ratio is similar to the values for planetary nebulae. It is possible that the chemical compositions in active symbiotic stars can be influenced by the active component. Tatarnikova et al. \cite{Tatar2011} suggested this possibility for the symbiotic Nova PU Vul. The electron number density is higher in quiescence than in the active state. This is apparently related to an increase of the ionizing radiation flux and subsequent excitation of forbidden lines in more distant rarefied regions of the gaseous envelope surrounding the system.

\section{Cool component}
The spectral type of the cool component can be estimated from the molecular bands of [TiO], applying the spectral indices introduced by Kenyon and Fernandez-Castro \cite{KenFern87}. The derived indices are presented in Table 4. The spectral indices increase (spectral type becomes redder) when the brightness of the system decreases. Our model SED shows that this behavior of the spectral indices defining the depths of the [TiO] molecular bands is due to the fact that the red part of the spectrum is not determined by the cool component alone, and includes the contribution of an additional fairly cool (T$_eff$~$\approx~$6000--10~000 K) radiation source associated with the hot component (the pseudophotosphere).
 
However, in quiescence, i.e., in the brightness minima in 2007 (on September 27, 2007) and 2009 (on June 1, 2009), the spectral type of the cool component corresponded to the type determined for Z~And by Murset and Schmid \cite{MS}. We also determined the spectral type of the cool component by comparing the observed and calculated SEDs, presented below. In this case, the spectral type of the cool component is also consistent with the type derived by us from the [TiO] spectral lines in quiescence. Table 4 shows that the spectral types determined from spectral lines with shorter wavelengths are earlier than thos derived from lines with larger wavelengths. This pattern was also noted by Kenyon and Fernandez-Castro \cite{KenFern87} and Tatarnikova et al. \cite{Tatar2009}. This difference is related to the large contribution of sources other than the giant to the emission in the 6215~\AA\, TiO molecular band.

\section{Model SED}
We represented the continuum SED as the sum of the contributions of the cool component, gaseous nebula, and relatively cool pseudophotosphere. Thus, the total continuum flux in our model is F($\lambda$)=F$_g$($\lambda$)+F$_p$($\lambda$)+F$_n$($\lambda$), where F$_g$($\lambda$) is the flux from the giant, F$_p$($\lambda$) the flux from the pseudophotosphere, and F$_n$($\lambda$) the flux from gaseous hydrogen nebula. The total flux F($\lambda$) was fitted to the observed flux using the $\chi^{2}$ criterion

\begin{eqnarray}
\chi^{2}=\sum_{i=1}^{N}\left[\frac{F_{obs}(\lambda_i)-F(\lambda_i)}
{\Delta~F_{obs}(\lambda_i)}\right]^2,
\end{eqnarray}

where F$_{obs}$($\lambda_i$) is the observed continuum flux, F($\lambda_i$)  -- the theoretical flux, $\Delta$~F$_{obs}$($\lambda_i$) -- the errors in the observed fluxes, and N~$\approx$~1700 -- 1800  the number of fluxes taken from the continuous spectrum. We excluded from our set of continuum fluxes emission lines and the range from 3645 to 3745~\AA, where large numbers of Balmer lines form blends and the continuum cannot be determined accurately. The errors in the fluxes were about 10\%. The method used is described in detail in \cite{S}. 

Figures 11--19  show the model and observed spectra of Z And obtained from 2008 to 2010. 
In the model, we used the synthetic photospheric spectrum of an M-giant \cite{FPT}, which we used to determine the spectral type of the cool component. The emission of the nebula was approximated as the emission of a hydrogen plasma, including the contributions of free--bound and bound--bound transitions. In this way, we derived the average electron temperature (T$_e$) and emission measure (EM) of the nebula. In our model, we also distinguished the contribution of the relatively cool pseudophotosphere of the hot component. This component was separated out from the spectrum and compared to synthetic spectra calculated by Munari et al. \cite{MSC2005} using Kurucz models. For temperatures above 8000 K, we used a black-body approximation. We derived the temperature (T$_p$), effective radius (R$_p$), and luminosity (L$_p$) of the pseudophotosphere of the hot component. The corresponding parameters are presented in Table 5.

\begin{center}
Fig 11. Observed spectra of Z~And corrected for interstellar reddening and the computed continuum~SED.
\end{center}

\begin{center}
Fig 12. Same as on Fig.~11.
\end{center}

\begin{center}
Fig 13. Same as on Fig.~11.
\end{center}

\begin{center}
Fig 14. Same as on Fig.~11.
\end{center}

\begin{center}
Fig 15. Same as on Fig.~11.
\end{center}

\begin{center}
Fig 16. Same as on Fig.~11.
\end{center}

\begin{center}
Fig 17. Same as on Fig.~11.
\end{center}

\begin{center}
Fig 18. Same as on Fig.~11.
\end{center}

\begin{center}
Fig 19. Same as on Fig.~11.
\end{center}

Table 5 also lists the physical parameters of the giant and gaseous nebula used for our model SED. We list the spectral type (Sp) and effective temperature (T$_{eff}$) of the giant, and the average electron temperature (T$_e$) and emission measure (EM) of the emission of the gaseous nebula. The last column gives the value of $\chi^{2}$$_{red}$,  normalized to the number of degrees of freedom n: $\chi^{2}$$_{red}$=$\chi^{2}$/n. The value of n was close to be 1700, determined as the number of selected continuum points N minus the number of free parameters. In our model SED, we used six free parameters: the spectral type (Sp) and effective temperature (T$_{eff}$) of the cool component, temperature (T$_p$) and radius (R$_p$) of the pseudophotosphere, and electron temperature (T$_e$) and emission measure (EM) of the nebula plasma. 

Our model computations show that, when the U brightness of Z And decreased (October 28, 2008 -- August 28, 2009, Figs. 11 and 12), the contribution of the cool pseudophotosphere was relatively small. However, when the U brightness increased (U<10$^m$), the pseudophotosphere dominated in the visual (Figs. 15-19, model spectra from August 25, 2009 to November 18, 2010), while the contribution of the nebular component to the continuum was weak. There were difficulties in detecting the nebula using our method at the brightness maxima, while the nebular emission increased as the stellar brightness decreased in the spectra taken from June 5, 2010 to December 18, 2010. Our observations indicate dramatic changes in the continuum SED during the brightness increase from August 28, 2009 (U$\approx$9.9) to September 25, 2009 (U$\approx$9.0). The nebular continuum decreased by a factor of 20, the [NeV] and [FeVII] lines disappeared, and the luminosity of the relatively cool pseudophotosphere increased by a factor of 11 (Table 5). This provides evidence for the formation of a powerful pseudophotosphere over several weeks, i.e., in a very short time. The simultaneous presence of a cool pseudophotosphere and nebular emission suggests a disk-like structure for the pseudophotosphere of the hot source. Such a model was first suggested by Skopal \cite{S}, based on amodel SED for active symbiotic stars \cite{SPB}.

\section{Conclusion}
We have performed an analysis of the symbiotic system Z And based on spectrophotometric observations obtained from 2006 to 2010, during which time there were three episodes of brightness increases. The dependence of the fluxes of lines of elements with various ionization potentials on the activity of the system shows that lines with low ionization potentials are correlated and lines with high ionization potentials anti-correlated with the light curve. However, a considerable decrease of the fluxes in all lines was observed at phases close to maximum brightness. In our opinion, this decrease in the observed line fluxes is related to an increased contribution of a pseudophotosphere to the total radiation, so that the line fluxes are veiled by the flux in the continuum. Our model SED shows that the contribution of the pseudophotosphere dominates in the spectrum in this interval, when its luminosity was the highest over the total period covered by the observations. 

The Raman scattering line, which provides information about the region close to the white dwarf, was observed during the brightness minimum in 2007 only, not in the minimum in 2009, despite the high temperature of the white dwarf at the latter epoch and the fact that white dwarf was not eclipsed by the giant at that phase. We suggest that the absence of this line is related to screening of the region close to the white dwarf by a disk-like object, which we call the  pseudophotosphere. 

We have considered variations of the hot component, giant and gaseous nebula in the active state and in quiescence. We have shown that the temperature of the hot component decreases to 90 000 K when the system brightness increases, and increases to 150 000--170~000~K when the brightness decreases. The electron density in the gaseous nebula in the region of formation of the helium lines is N$_e$~$\approx$~10$^{10}$ -- 10$^{12}$~cm$^{-3}$. A certain tendency for the gas density to decrease with decreasing brightness of the system is observed. On the other hand, the ratios of the fluxes in the helium lines shift into the region of higher densities at epochs of maximum brightness. We estimated the electron density and temperature of the extended regions of the gaseous nebula using the ratio of the [OIII] and [NeIII] nebular lines. The temperature of the nebula was 12 000--20 000 K; its electron density was N$_e$~$\approx$~10$^{7}$~cm$^{-3}$ in quiescence, and N$_e$~$\approx$~10$^{6}$~cm$^{-3}$ in the active state. The theoretical and observed values of R$_1$ and R$_2$ agree better when the O/Ne abundance ratio is the same as for planetary nebulae. It is possible that the active components may influence the chemical compositions of active symbiotic stars. 

The spectral index of the cool component in the quiescent states in both 2007 and 2009 corresponded to the value derived by Murset and Schmid \cite{MS}. In the active state, the system became less red. As our model SED shows, this seems to be related to the fact that a significant contribution to the red spectrum in the active state is provided by an additional radiation source~--~the pseudophotosphere of the hot component. 

Our model SED calculations confirm that a relatively cool (5250--11 500 K) pseudophotosphere was present around the hot component of Z And in 2008 and 2009--2010; the radius and temperature of the pseudophotosphere increased with increasing brightness of the system. For example, the luminosity of the pseudophotosphere increased by a factor of 11 from August 28, 2009 to September 25, 2009, and dominated in the visual. This shows that a powerful pseudophotosphere can form very rapidly, over several weeks. The low temperature of the pseudophotosphere, simultaneous presence of strong nebular emission, and absence of the Raman scattering line, in spite of the high temperature of the hot source (150~000--170~000~K), suggest that this pseudophotosphere had a disk-like structure. 

The presence of collimated ejections in the system demonstrate the existence of such an object. We observed collimated ejections, or jets, as features at the blue and red sides of the H$_{\alpha}$ and H$_{\beta}$ line wings. We found the radial velocity of the matter in the jets. In 2009, during the brightness increase toward the maximum, this velocity was similar to the value in 2006~--~close to  1200 km/s. However, after the star reached the brightness maximum in 2009, new components arose with substantially higher velocities, close to 1800 km/s.

\section{Acknowledgments}
This study was supported by the Ukrainian State Fund for Basic Research (grant F28.2/081) and the Slovak Academy of Sciences (grant No. 2/0038/10).

\newpage
\begin{table}[!ht]
\caption{Log of spectroscopic observations of Z And}
\bigskip
\begin{tabular}{lccc}
\hline

Date&	JD (2450000+) &Phase & Range \\
\hline
16.09.06&3995.455&0.931&3750-7600\AA\\
10.10.06&4019.446&0.962&3876-7574\AA\\
14.01.07&4115.253&0.087&3750-7513\AA\\
28.07.07&4310.382&0.346&4350-5300\AA\\
27.09.07&4371.481&0.426&3774-7575\AA\\
07.07.08&4654.535&0.800&3776-7574\AA\\
07.07.08&4655.405&0.800&3775-7949\AA\\
10.08.08&4689.437&0.845&3750-7574\AA\\
15.08.08&4694.469&0.851&4075-5050\AA\\
25.09.08&4734.583&0.905&3749-7574\AA\\
12.10.08&4752.208&0.928&4074-5075\AA\\
23.10.08&4763.391&0.943&3349-7163\AA\\
03.11.08&4774.358&0.957&4075-5100\AA\\
07.11.08&4778.343&0.962&3750-7513\AA\\
01.06.09&4983.524&0.233&3324-7375\AA\\
14.07.09&5026.522&0.290&3325-6974\AA\\
28.08.09&5071.562&0.349&3349-7416\AA\\
25.09.09&5099.551&0.386&3350-7425\AA\\
26.09.09&5100.535&0.387&3349-7400\AA\\
29.09.09&5104.226&0.392&4074-5050\AA\\
21.05.10&5337.499&0.700&3525-7574\AA\\
04.07.10&5382.465&0.759&4125-5075\AA\\
05.07.10&5383.437&0.760&3525-7600\AA\\
16.08.10&5425.315&0.815&4125-5075\AA\\
16.08.10&5425.466&0.815&3405-7575\AA\\
15.09.10&5454.506&0.853&3300-7575\AA\\
18.11.10&5519.293&0.939&3300-7575\AA\\
18.11.10&5519.293&0.939&4125-5075\AA\\
\hline
\end{tabular}
\end{table}

\newpage
\begin{table}[!ht]
\caption{Emission-lines fluxes of Z And in units
(10$^{-12}$~erg~cm$^{-2}$~s$^{-1}$)}
\bigskip
\begin{tabular}{lp{6mm}p{6mm}p{6mm}p{6mm}p{6mm}p{6mm}p{6mm}p{6mm}p{6mm}p{6mm}p{6mm}p{6mm}p{6mm}}
\hline
JD&3995&4019&4115&4310&4371&4654&4689&4694&4734&4752&4763&4774&4778\\
Phase&0.93&0.96&0.09&0.35&0.43&0.80&0.85&0.85&0.91&0.93&0.94&0.96&0.96\\
\hline
3426 [NeV]&&&&&&&&&&&11.2&&\\
H$_{9}$ &&&4.9&&5.2&3.6&3.3&&2.9&&2.9&&3.9\\
3869 [NeIII]&4.6&5.1&2.8&&1.2&5.3&5.8&&5.8&&5.7&&6.6\\
H$_{8}$ &&&7.7&&9.0&9.3&8.6&&6.3&&6.6&&5.8\\
3968 [NeIII]+& & & && & & && && && \\
H$_{\epsilon}$ &12.9&13.2&10.9&&9.5&11.4&10.9&&8.5&&8.7&&9.8\\4026 HeI&2.2&2.2&2.1&&1.2&3.9&3.2&&2.8&&2.4&&2.4\\
H$_{\delta}$ &15.9&15.2&16.6&&14.5&14.0&13.0&16.3&12.5&&12.8&13.4&13.2\\
H$_{\gamma}$ &22.9&23.1&21.6&&24.4&25.1&23.3&23.7&19.2&19.9&17.8&17.4&18.7\\
4363 [OIII]&3.3&5.6&3.3&1.2&1.3&5.8&5.9&6.2&6.3&6.3&5.5&5.4&5.4\\
4388 HeI&3.2&3.7&2.3&0.8&0.9&3.7&3.1&3.0&2.5&2.6&2.6&2.5&2.6\\
4471 HeI&4.6&3.9&3.3&1.45&2.5&5.0&4.8&4.9&4.2&4.1&3.8&3.6&3.4\\
%4541 HeII&--&--&1.8&1.8&2.3&1.6&1.4&1.5&1.5&1.5&1.5&1.5&1.5\\
4640 NIII+&&&&&&&&&&&&&\\
CII&13.3&18.5&13.7&5.7&5.3&11.7&11.2&13.2&14.5&13.4&13.9&13.4&13.3\\
4686 HeII&6.9&16.2&32.9&35.7&32.3&17.5&18.0&28.2&27.6&27.8&28.6&28.4&28.3\\
H$_{\beta}$ &80.8&81.4&61.6&64.5&66.8&79.0&65.7&64.2&57.6&60.4&57.0&54.4&54.7\\
4922 HeI&7.1&8.1&6.8&2.2&4.0&5.8&5.4&5.0&5.0&4.7&4.8&4.5&5.2\\
5007 [OIII]&8.7&12.0&5.7&0.9&0.8&8.8&10.0&10.3&10.6&10.0&9.5&8.0&8.1\\
5016 HeI&6.3&5.8&5.6&2.0&3.2&7.3&5.6&6.6&5.3&5.3&5.0&4.6&4.7\\
5412 HeII&+& +&2.8&&2.7&1.4&1.2&&1.9&&1.5&&1.8\\
5721 [FeVII]&--&--&0.66&&2.55&0.6&0.6&& 0.7&&0.7&&0.8\\
5876 HeI&13.8&12.9&11.1&&9.5&11.4&13.7&&12.6&&11.9&&12.0\\
6087 [FeVII]&--&--&1.1&&3.8&0.9&1.1&&1.3&&1.3&&1.4\\
H$_{\alpha}$ &308&316&293&&341&345&310&&265&&260&&240\\
6678 HeI&9.8&11.1&10.9&&7.2&7.1&8.0&&6.6&&6.0&&6.5\\
6830&--&--&--&&10.2&--&--&&--&&--&&\\
7065 HeI 7065&9.3&9.5&10.4&&7.8&6.9&8.7&&8.4&&&&7.8\\
\hline
\end{tabular}
\end{table}

\newpage
\begin{table}[!ht]
\centerline{Table 2. Contd.}
\bigskip
\begin{tabular}{lcccccccccc}
\hline
JD&4983&5026&5071&5099/100&5104&5338&5383/84&5425/26&5454&5519\\
Phase&0.23& 0.29&0.35&0.39&0.39&0.70&0.76&0.82&0.87&0.93\\
\hline
3426 [NeV]      &27.0&19.7&20.5&-- &&--  &--  &    &5.3 &8.0\\
3586 [FeVII]    &2.8 &2.6 &2.7 &-- &&--  &--  &--  &--  &--\\
3760 [FeVII]    &5.0 &5.0 &4.9 &-- &&--  &--  &+   &+   &+\\
3869 [NeIII]    &4.1 &3.2 &2.7 &3.5&&4.6 &5.1 &7.1 &6.3 &6.1\\
H$_{8}$         &5.3 &5.9 &5.1 &3.4&&9.9 &9.2 &8.2 &6.4 &6.1\\
3968 [NeIII]+H$_{\epsilon}$  &7.0 &7.8 &10.0&6.2&&13.2&15.7&16.7&11.2&10.8\\
HeI 4026        &1.1 &1.4 &2.0 &2.0&&2.8 &3.3 &2.7 &2.2 &1.4\\
H$_{\delta}$    &10.0&11.0&13.9&8.3&5.8&21.3&23.2&18.9&15.4&15.1\\
H$_{\gamma}$    &14.4&16.4&19.8&11.2&9.4&32.9&34.3&28.7&20.8&18.5\\
4363 [OIII]     &3.7&3.0&2.9&2.4&2.1&4.0&6.7&6.9&5.3&5.1\\
4388 HeI        &1.3&1.7&2.8&1.8&1.8&4.0&4.3&3.3&2.5&3.0\\
4471 HeI        &1.6&1.7&2.1&2.0&2.2&4.3&5.1&4.2&2.7&1.5\\
%4541 HeII       &1.9&1.6&1.7&0.5&0.6&+&1.3&1.5&+&+\\
4640 NIII+CII   &10.0&10.0&13.2&7.8&7.1&15.0&18.7&17.0&14.1&13.5\\
4686 HeII       &32.9&30.3&37.2&7.9&7.5&9.3&21.0&24.5&19.0&23.7\\
H$_{\beta}$     &43.4&55.0&57.0&33.5&32.9&89.6&94.7&80.9&59.4&48.2\\
4922 HeI        &2.9&5.1&6.6&3.0&2.8&7.1&8.6&7.4&5.2&5.6\\
5007 [OIII]     &3.1&3.0&3.0&8.3&7.5&8.6&13.6&13.9&11.4&9.0\\
5016 HeI        &1.9&2.9&4.0&3.2&3.8&8.3&9.3&8.1&6.4&5.1\\
5412 HeII       &2.6&2.2&2.7&1.0&&+&1.3&1.4&1.6&1.7\\
5721 [FeVII]    &2.5&2.1&1.8&--&&--&--&+&+&+\\
5876 HeI        &5.2&6.1&7.0&7.0&&14.9&17.7&17.5&14.3&9.1\\
6087 [FeVII]    &3.8&3.4&2.6&0.4&&--&--&+&+& +\\
H$_{\alpha}$    &222&272&259&164&&407&424&438&306&229\\
6678 HeI        &6.0&9.4&11.0&4.5&&9.2&11.2&11.6&8.8&8.1\\
7065 HeI        &4.3&&6.6&2.5&&8.7&13.3&13.0&11.1&7.3\\
\hline
\end{tabular}
\end{table}

\newpage
\begin{table}[!ht]
%\setcaptionmargin{0mm}
%\onelinecaptionstrue
%\captionstyle{flushleft}
\caption{ Temperature of the hot component of Z And}
\bigskip
\begin{tabular}{cccc}
\hline
Date&JD(2450000+)&Phase&Temperature, K\\
\hline

16.09.06&3995.455&0.93&90000\\
10.10.06&4019.446&0.96&112000\\
14.01.07&4115.253&0.09&149000\\
28.07.07&4310.382&0.35&154000\\
27.09.07&4371.481&0.43&146000\\
07.07.08&4654.535&0.80&113000\\
10.08.08&4689.437&0.85&120000\\
15.08.08&4694.469&0.85&138000\\
25.09.08&4734.583&0.91&142000\\
12.10.08&4752.208&0.93&140000\\
23.10.08&4763.391&0.94&144000\\
03.11.08&4774.358&0.96&146000\\
07.11.08&4778.343&0.96&146000\\
01.06.09&4983.524&0.23&170000\\
14.07.09&5026.522&0.29&153000\\
28.08.09&5071.562&0.35&161000\\
25.09.09&5099.551&0.39&118000\\
26.09.09&5100.535&0.39&113000\\
29.09.09&5104.226&0.39&114000\\
21.05.10&5337.499&0.70&94500\\
04.07.10&5382.465&0.76&115000\\
16.08.10&5425.413&0.82&129000\\
15.09.10&5454.506&0.87&127000\\
18.11.10&5519.395&0.94&147000\\
\hline
\end{tabular}
\end{table}

\newpage
\begin{table}[!ht]
%\setcaptionmargin{0mm}
%\onelinecaptionsfalse
%\captionstyle{flushleft}
\caption{Spectral indices [TiO]$_1$ and [TiO]$_2$ of the [TiO]~
$\lambda$~6215\AA~ and $\lambda$~7125\AA\, molecular bands and spectral type of the cool component}
\bigskip
\begin{tabular}{ccccccc}
\hline
Date&JD(2450000+)&Phase&[TiO]$_1$&ST$_1$&[TiO]$_2$&ST$_2$\\
\hline

16.09.06&3995.455&0.931&0.22&M0.5&0.53&M2.8\\
10.10.06&4019.446&0.962&0.24&M0.7&0.54&M2.8\\
14.01.07&4115.253&0.087&0.27&M0.9&0.61&M3.3\\
27.09.07&4371.481&0.426&0.51&M3.0&0.89&M4.5\\
07.07.08&4655.405&0.800&0.32&M1.4&0.50&M2.7\\
10.08.08&4689.437&0.845&0.34&M1.5&0.65&M3.5\\
25.09.08&4734.583&0.905&0.38&M1.9&0.61&M3.3\\
23.10.08&4763.391&0.943&0.36&M1.7&--&--\\
07.11.08&4778.343&0.962&0.41&M2.4&0.77&M4.0\\
01.06.09&4983.524&0.233&0.54&M3.3&0.96&M4.6\\
14.07.09&5026.522&0.290&0.49&M2.8&--&--\\
28.08.09&5071.562&0.349&0.39&M2.0&0.58&M2.8\\
25.09.09&5099.551&0.386&0.24&M0.7&0.43&M2.0\\
26.09.09&5100.535&0.387&0.23&M0.5&0.43&M2.0\\
21.05.10&5337.499&0.700&0.21&M0.3&0.33&M1.4\\
05.07.10&5383.437&0.760&0.26&M0.8&0.33&M1.4\\
16.08.10&5425.466&0.815&0.29&M1.1&0.48&M2.6\\
15.09.10&5454.506&0.853&0.37&M1.7&0.52&M2.7\\
18.11.10&5519.395&0.939&0.22&M0.5&0.53&M2.8\\
\hline
\end{tabular}
\end{table}

\newpage
\begin{table}[!ht]
%\setcaptionmargin{0mm}
%\onelinecaptionsfalse
%\captionstyle{flushleft}
\caption{Physical parameters of the composite spectrum of Z And 
         derived by our SED-fitting analysis}
\bigskip
\begin{tabular}{ccccccccc}
\hline
Date&\multicolumn{2}{c}{Giant}&\multicolumn{3}{c}{Pseudophonosphere}&
\multicolumn{2}{c}{Nebula}&$\chi^2$$_{red}$\\
&ST&T$_{eff}$(K)&T$_p$(K)&R$_p$(R$_{\odot}$)&L$_p$(L$_{\odot}$)&T$_e$(K)&
EM(10$^{60}$cm$^{-3}$)\\
\hline
23.10.2008&4.0&3491&5750&11.5&130&32000&3.1&0.51\\
01.06.2009&4.3&3463&5250&7.9&43&41000&1.6&0.69 a)\\
14.07.2009&4.0&3491&5250&9.4&61&25000&1.3&0.55\\
28.08.2009&4.4&3453&5250&11.4&89&22000&2.3&0.31\\
25.09.2009&4.9&3405&9000&12.9&978&30000&0.1&0.31 b)\\
21.05.2010&4.2&3472&9500&12.0&1060&--&--&0.28 c)\\
05.07.2010&4.1&3481&9000&11.0&713&35000&1.7&0.39 d)\\
14.09.2010&4.8&3415&9500&8.7&556&27000&0.5&0.40\\
18.11.2010&4.9&3405&11500&6.8&729&22000&0.3&0.39\\
\hline\
\end{tabular}

\leftline{a) very faint stellar component}
\leftline{b) very faint nebula, large uncertainty in T$_e$ ($\pm$~10000 K)}
\leftline{c) nebular contrubution neglected}
\leftline{d) faint nebula appeared, the Balmer jump in emission can be recognized}
\end{table}

\end {document}
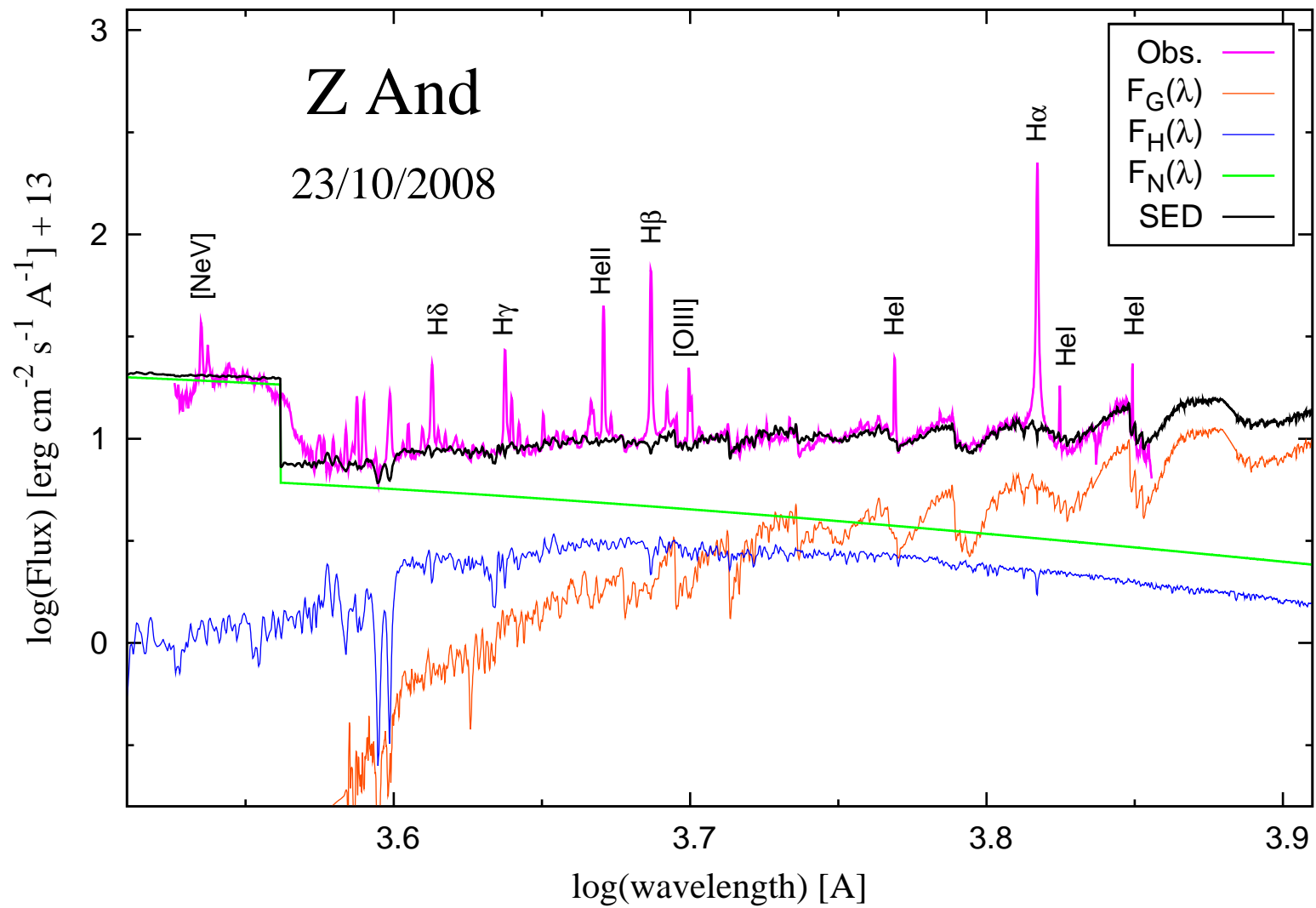
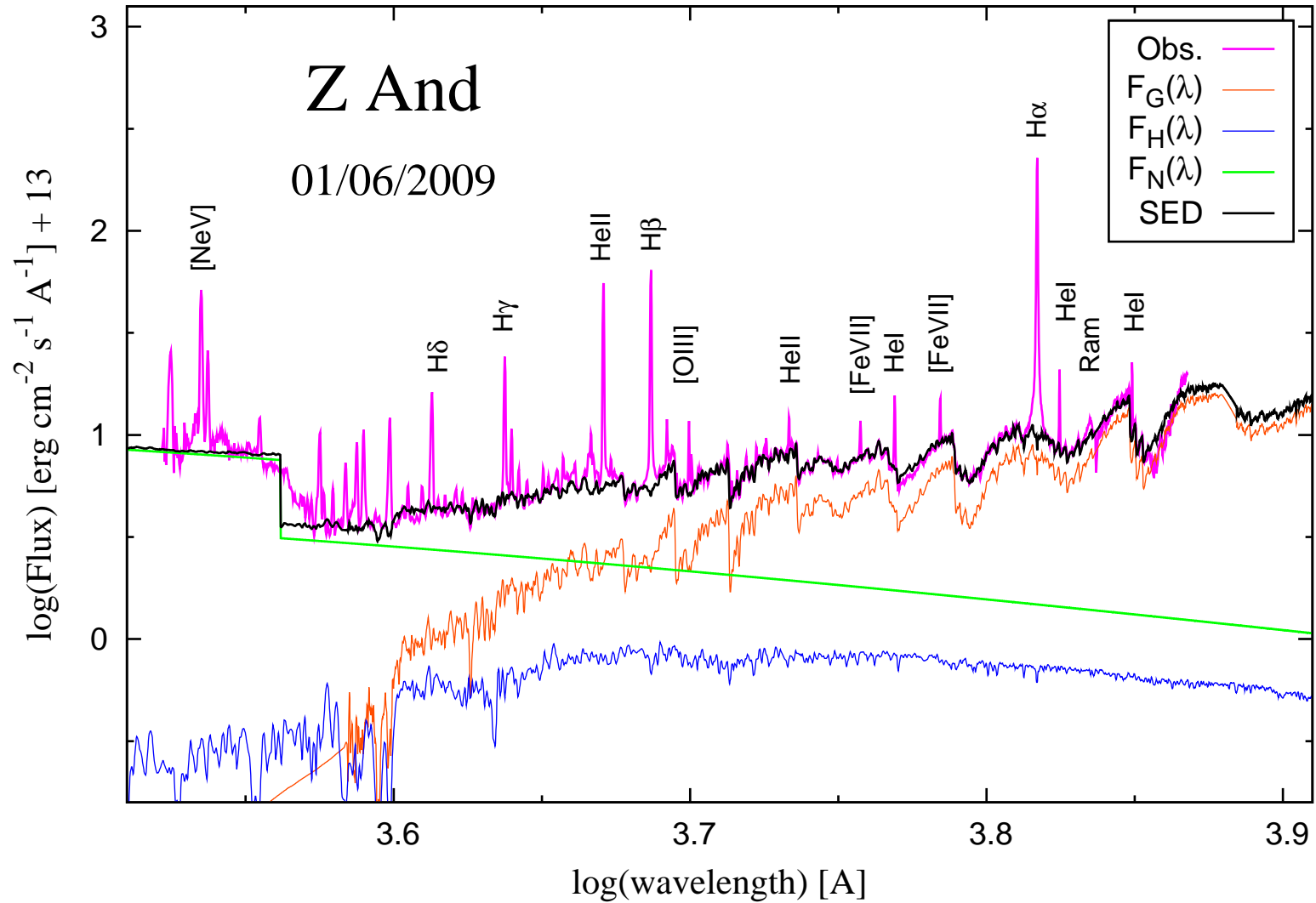
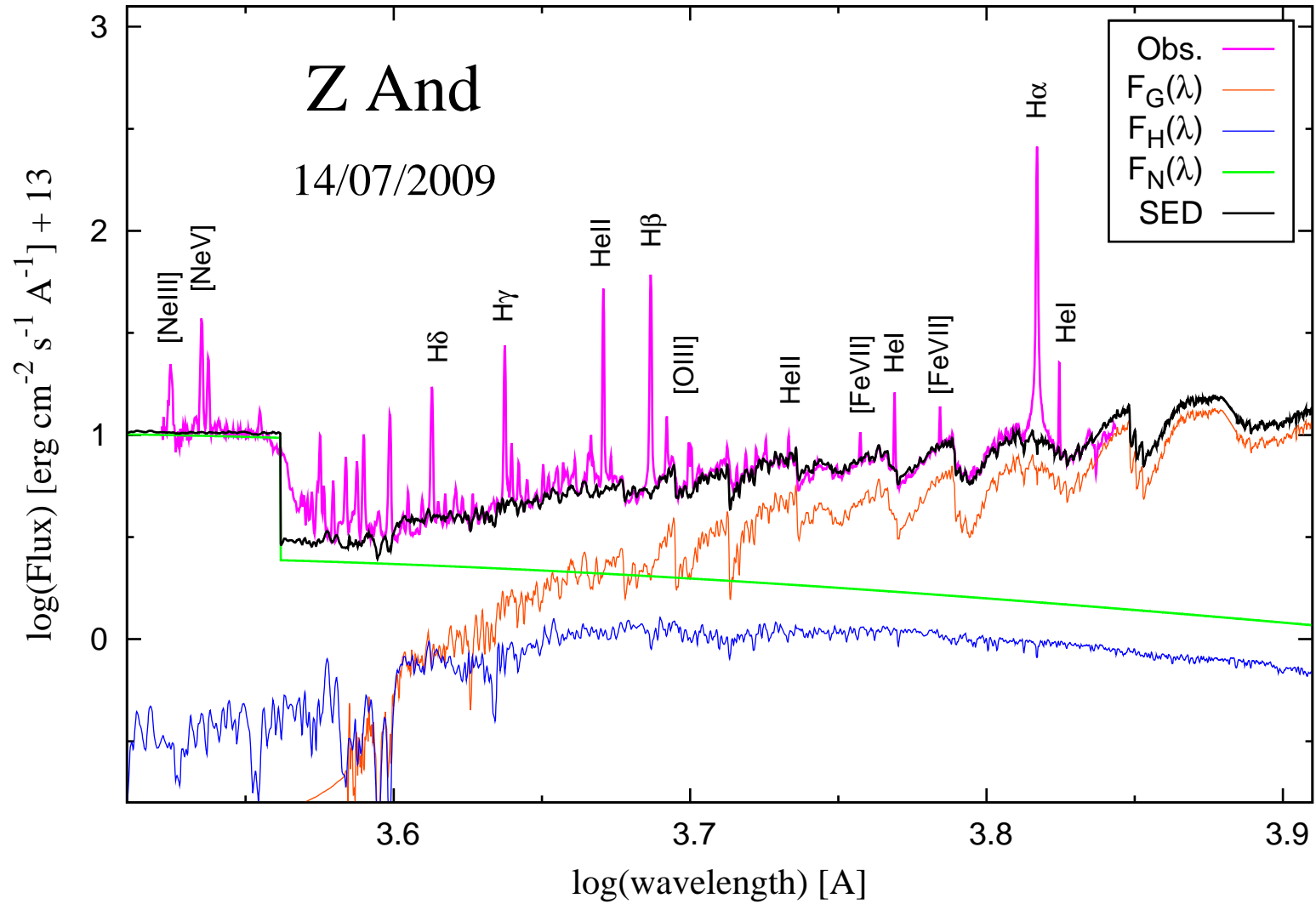
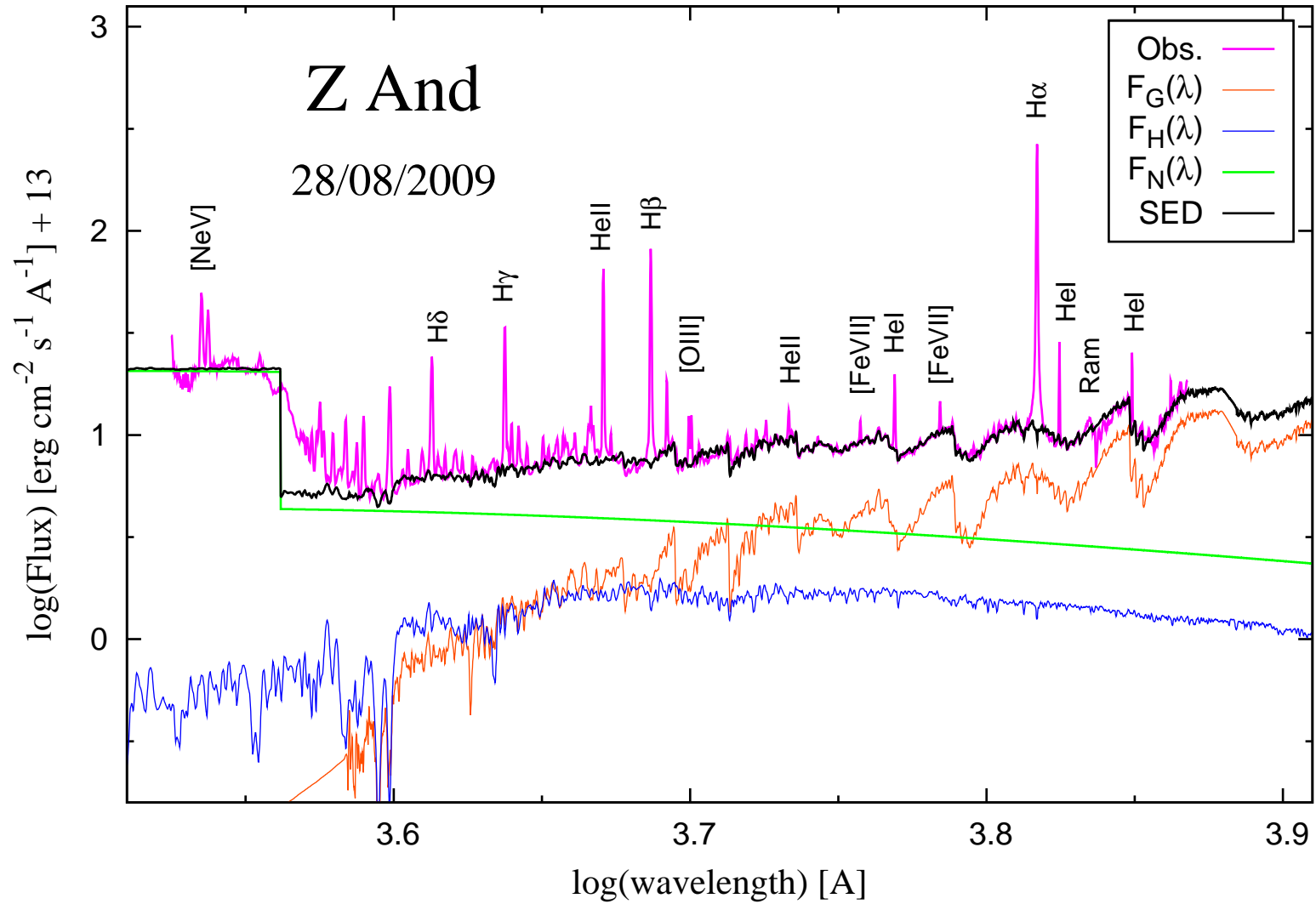
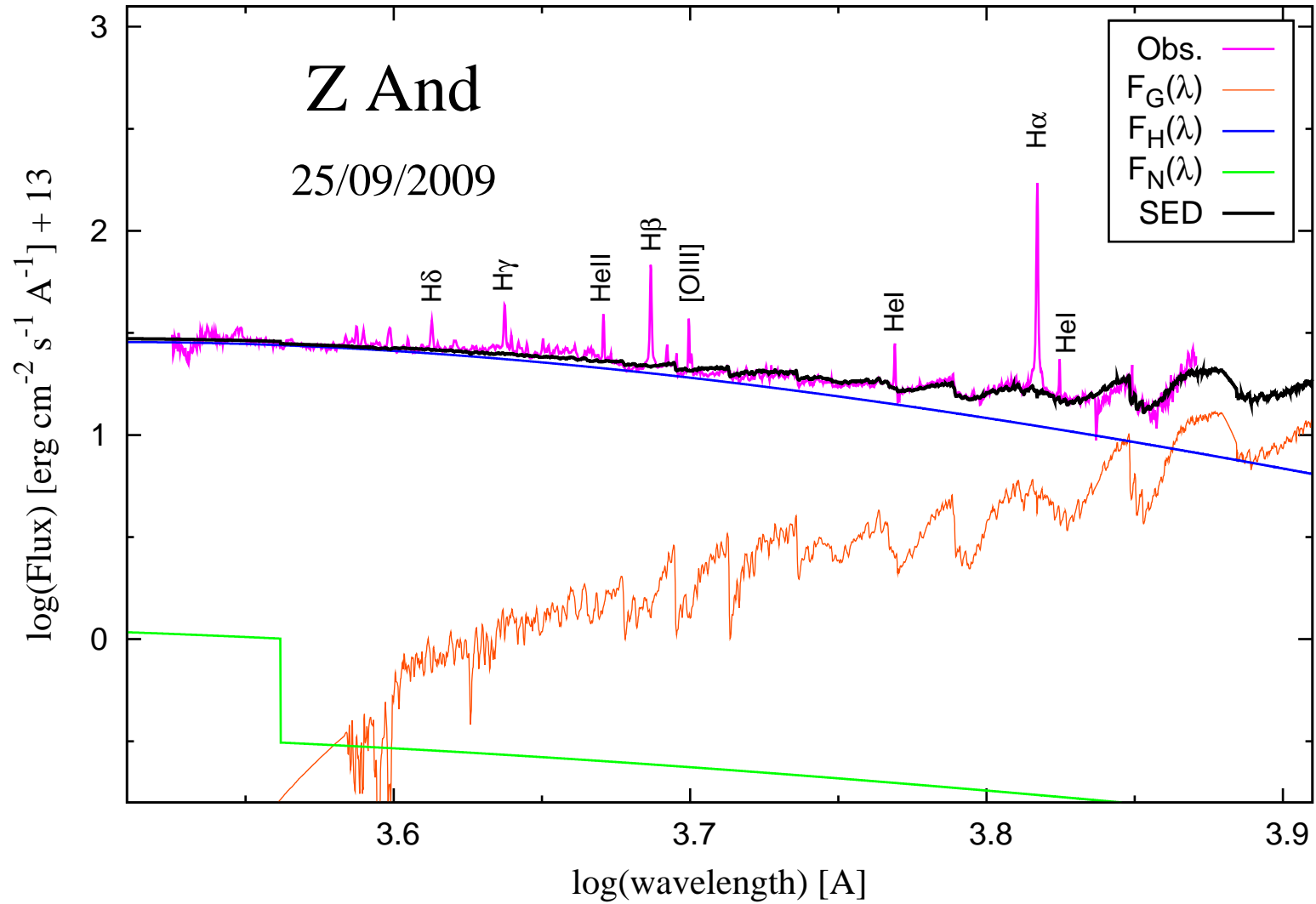
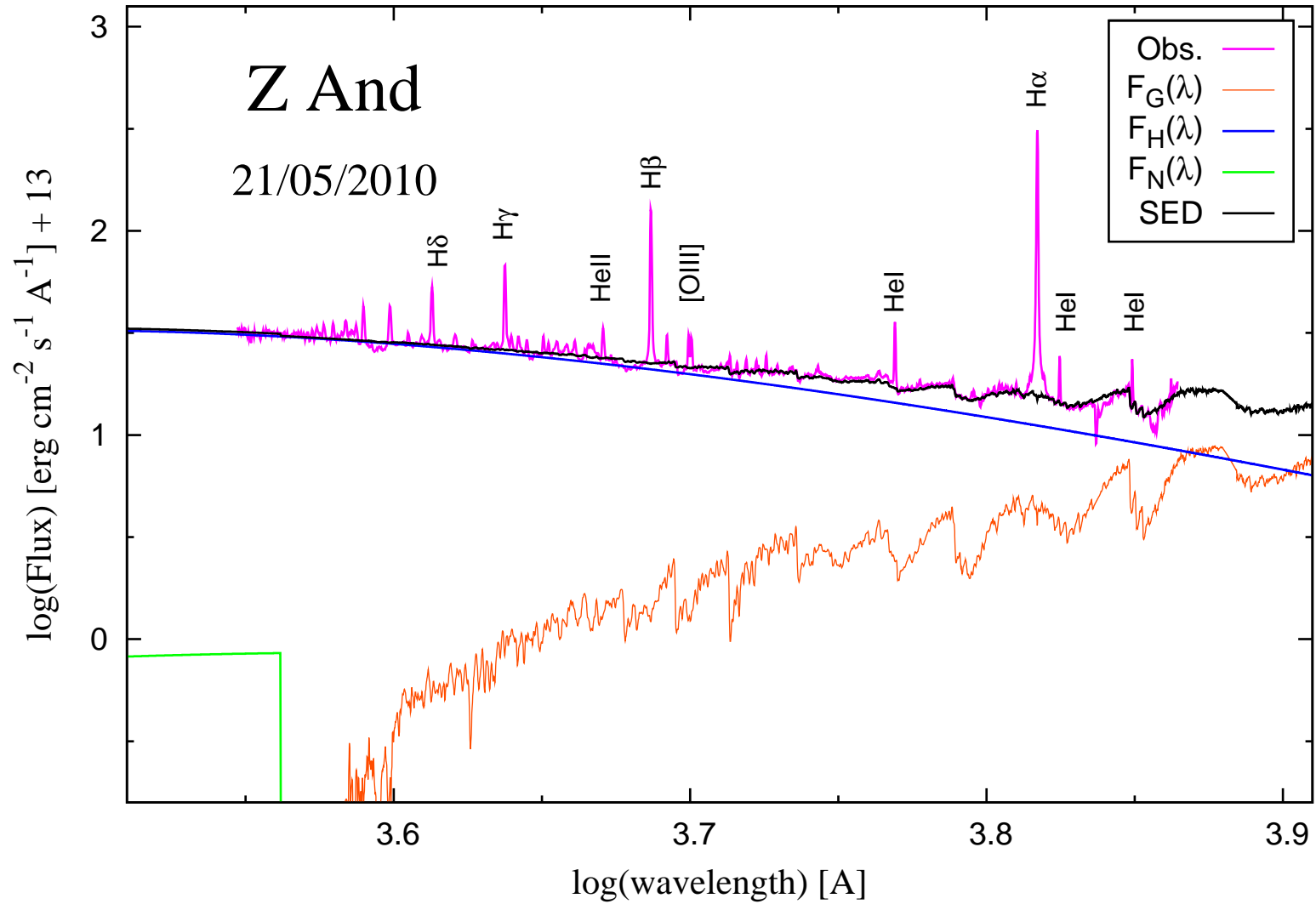
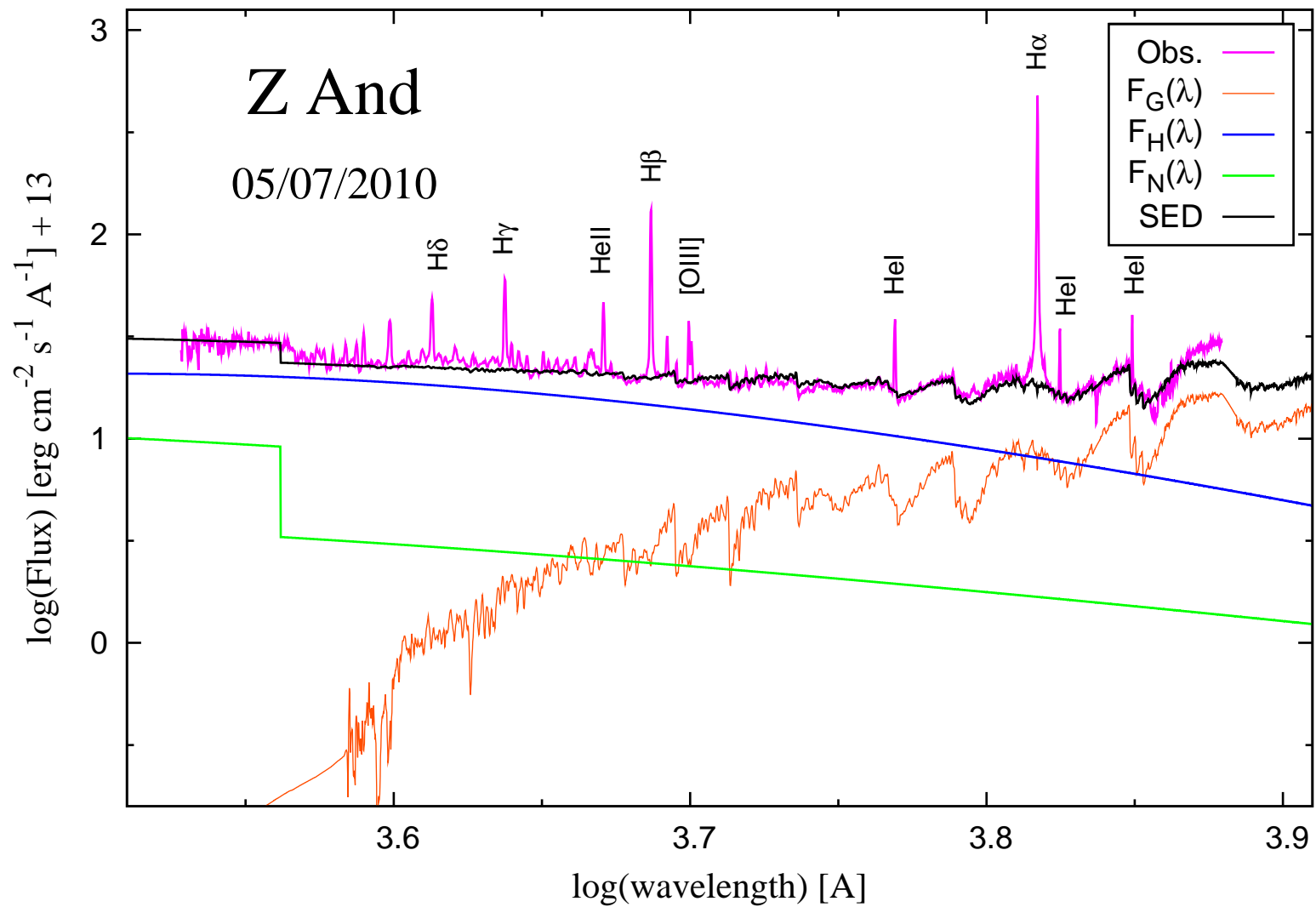
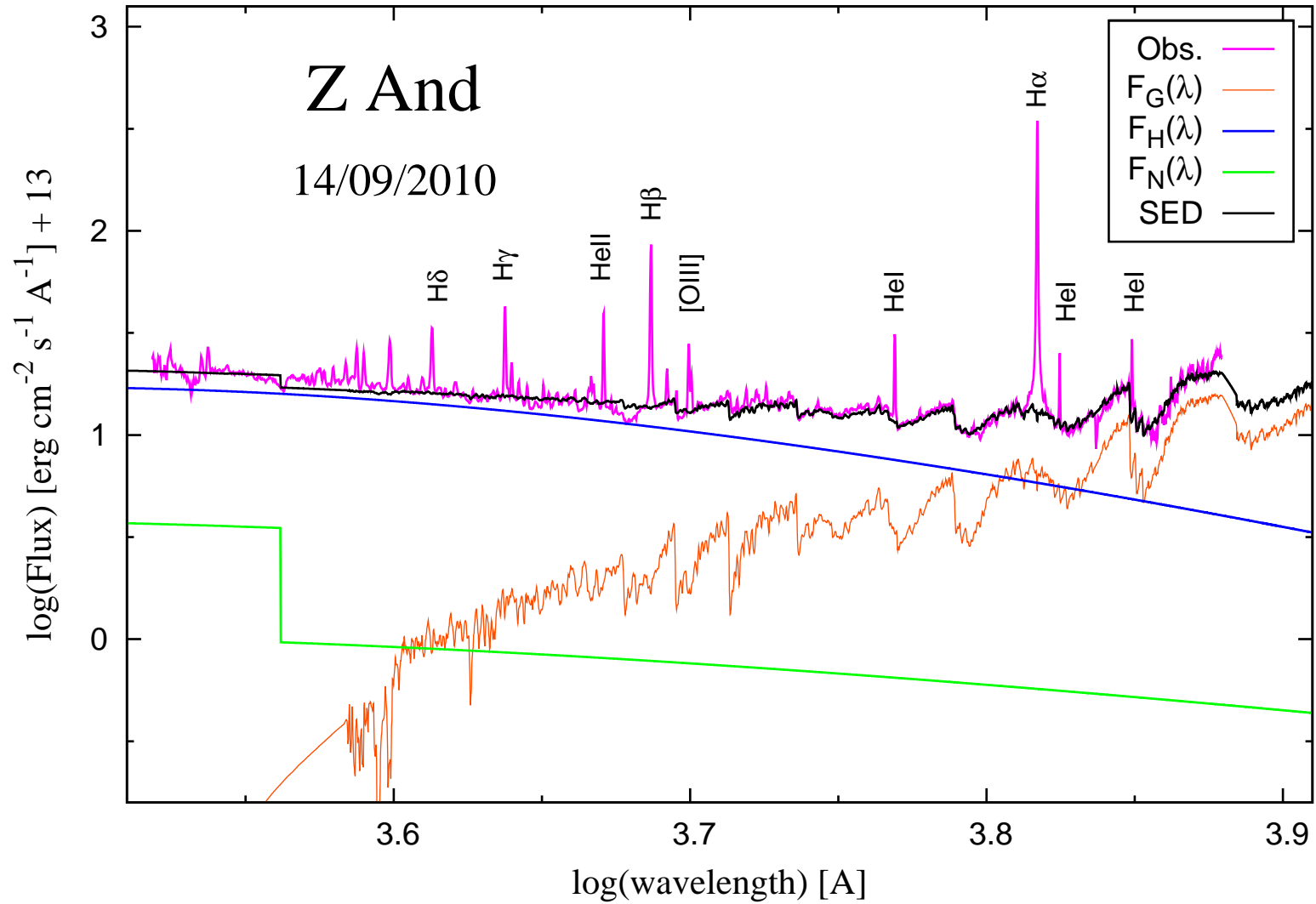
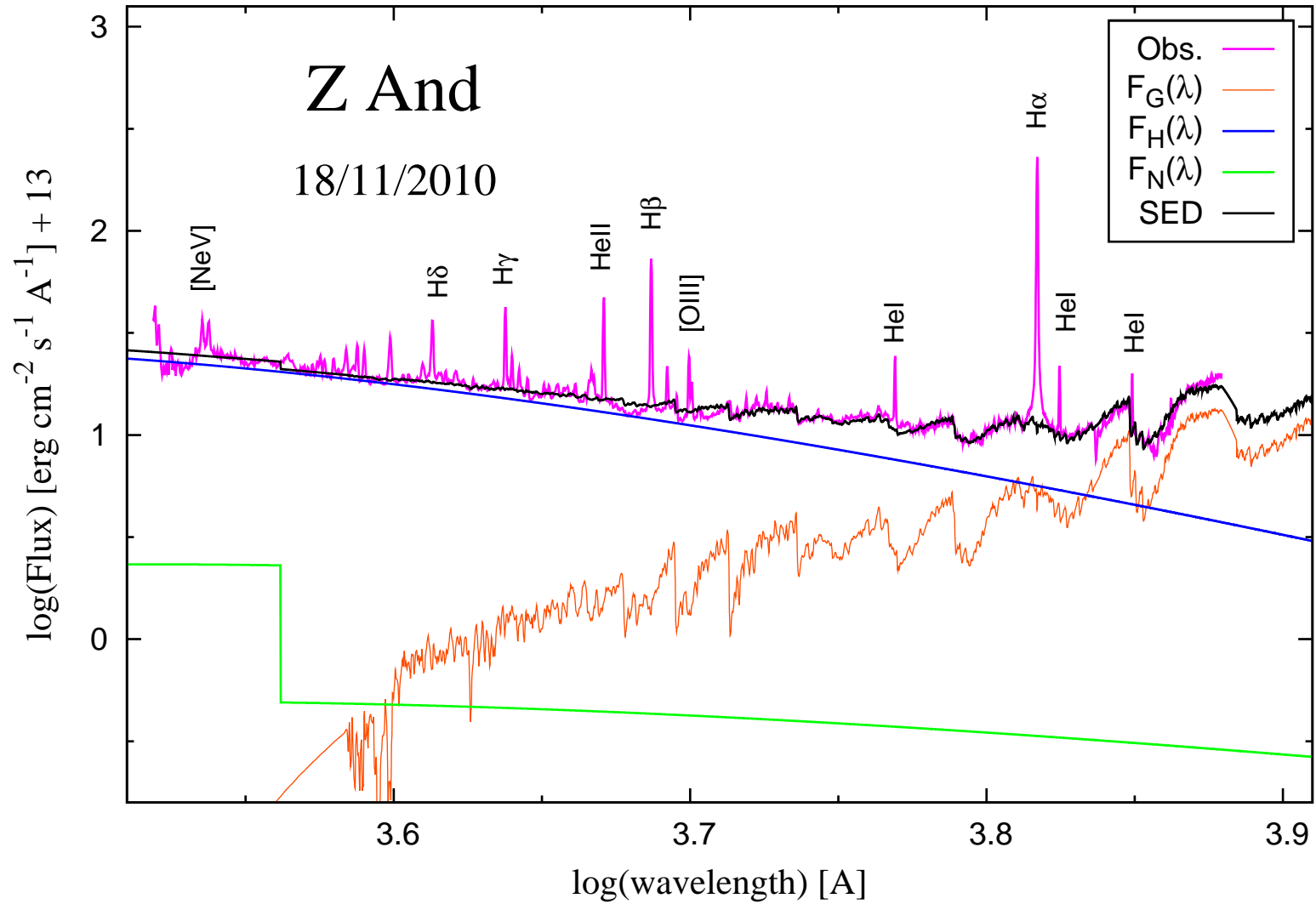
